\documentclass[aps,prb,a4paper,10pt,twocolumn,showpacs,floatfix,bibnotes,superscriptaddress,preprintnumbers,longbibliography]{revtex4-2}
\usepackage{float}
\usepackage{dcolumn,graphicx,color,booktabs,microtype,afterpage}
\usepackage{amssymb}
\usepackage{amsmath}
\usepackage[charter,greekuppercase=italicized]{mathdesign}
\usepackage{sidecap}
\usepackage[mathlines]{lineno}

\graphicspath{{./}{figure/}}
\renewcommand{\tablename}{Table}
\makeatletter\renewcommand{\fnum@figure}[1]{\figurename~\thefigure.~}\makeatother
\makeatletter\renewcommand{\fnum@table}[1]{\tablename~\thetable.}\makeatother

\newcount\hh \newcount\mm
\hh=\time \divide\hh by 60
\mm=\hh \multiply\mm by 60 \mm=-\mm
\advance\mm by \time
\def\now{\number\hh:\ifnum\mm<10{}0\fi\number\mm}

\usepackage[colorlinks,plainpages=false,linkcolor=blue,urlcolor=blue,citecolor=blue,pdfpagemode=UseNone,pdfstartview=FitBH]{hyperref}
\usepackage{nicefrac}

\newcommand{\tcr}[1]{\textcolor{black}{#1}}

\begin{document}

\makeatletter\renewcommand{\ps@plain}{%
\def\@evenhead{\hfill\itshape\rightmark}%
\def\@oddhead{\itshape\leftmark\hfill}%
\renewcommand{\@evenfoot}{\hfill\small{--~\thepage~--}\hfill}%
\renewcommand{\@oddfoot}{\hfill\small{--~\thepage~--}\hfill}%
}\makeatother\pagestyle{plain}

\preprint{\textit{Preprint: \today, \now}} %For internal use only, do not distribute.}}
%\linenumbers

\title{\tcr{Investigating the intrinsic anomalous Hall effect in MnPt$_3$ topological  semimetal}}
%\title{Intrinsic anomalous Hall effect in epitaxial MnPt$_3$ thin films}
%\tcr{Investigating the intrinsic anomalous Hall effect in MnPt$_3$ topological semimetal}
%\title{Intrinsic anomalous Hall effect in MnPt$_3$ topological semimetal}
%
\author{Jing Meng}
\affiliation{School of Physics, East China Normal University, Shanghai 200241, China}
\author{Hongru Wang}
% better to use new affiliation, not this one
\affiliation{Key Laboratory of Polar Materials and Devices (MOE), East China Normal University, Shanghai 200241, China}
\author{Kun Zheng}
\affiliation{School of Physics, East China Normal University, Shanghai 200241, China}
\author{Yuhao Wang}
\affiliation{School of Physics, East China Normal University, Shanghai 200241, China}
\author{Zheng Li}
\affiliation{School of Physics, East China Normal University, Shanghai 200241, China}
\author{Bocheng Yu}
\affiliation{School of Physics, East China Normal University, Shanghai 200241, China}
\author{Haoyu Lin}
\affiliation{School of Physics, East China Normal University, Shanghai 200241, China}
\author{Keqi Xia}
\affiliation{School of Physics, East China Normal University, Shanghai 200241, China}
\author{Jingzhong Luo}
\affiliation{School of Physics, East China Normal University, Shanghai 200241, China}
\author{Zengyao Wang}
\affiliation{School of Physics, East China Normal University, Shanghai 200241, China}
\author{Xiaoyan Zhu}
\affiliation{State Key Laboratory of Infrared Science and Technology, Shanghai Institute of Technical Physics, the Chinese Academy of Sciences, Shanghai 200083, China}
\author{Baiqing Lv}
\affiliation{Tsung-Dao Lee Institute, Zhangjiang Institute for Advanced Study, School of Physics and Astronomy, Shanghai Jiao Tong University, Shanghai 200240, China}
\author{Yaobo Huang}
\affiliation{Shanghai Synchrotron Radiation Facility, Shanghai Advanced Research Institute, Chinese Academy of Sciences, Shanghai 201204, China}
\author{Jie Ma}
\affiliation{Key Laboratory of Artificial Structures and Quantum Control, School of Physics and Astronomy, Shanghai Jiao Tong University, Shanghai 200240, China}
\author{Yang Xu}
\affiliation{School of Physics, East China Normal University, Shanghai 200241, China}
\affiliation{Key Laboratory of Polar Materials and Devices (MOE), East China Normal University, Shanghai 200241, China}
\author{Shijing Gong}
\affiliation{School of Physics, East China Normal University, Shanghai 200241, China}
\affiliation{Key Laboratory of Polar Materials and Devices (MOE), East China Normal University, Shanghai 200241, China}
\author{Tian Shang}\email[Corresponding authors:\\]{tshang@phy.ecnu.edu.cn}
\affiliation{School of Physics, East China Normal University, Shanghai 200241, China}
\affiliation{Key Laboratory of Polar Materials and Devices (MOE), East China Normal University, Shanghai 200241, China}
\author{Qingfeng Zhan}\email[Corresponding authors:\\]{qfzhan@phy.ecnu.edu.cn}
\affiliation{School of Physics, East China Normal University, Shanghai 200241, China}
\affiliation{Key Laboratory of Polar Materials and Devices (MOE), East China Normal University, Shanghai 200241, China}
\begin{abstract}
The cubic Cu$_3$Au-type $X$Pt$_3$ family ($X$ = V, Cr, and Mn) is a topological semimetal characterized by anti-crossing gapped nodal lines near the Fermi level, which give rise to significant Berry curvatures and thus to the anomalous Hall effect (AHE). Among the three members, CrPt$_3$ has been experimentally verified to exhibit a large anomalous Hall conductivity (AHC), while its counterparts MnPt$_3$ and VPt$_3$ remain largely unexplored. Here, a series of MnPt$_3$ thin films with varying thicknesses (20--70\,nm) was epitaxially grown on the MgO substrates using magnetron sputtering and was systematically investigated by magnetization, electrical resistivity, and Hall resistivity measurements. MnPt$_3$ films undergo a ferromagnetic transition at a Curie temperature $T_\mathrm{C}$, which increases as the film thickness increases, reaching $\sim$ 344\,K for the 70-nm-thick film. All the anomalous Hall transport properties of MnPt$_3$ films, including the resistivity, conductivity, and angle, exhibit a strong correlation with their magnetic properties. The scaling analysis suggests that the intrinsic Berry-curvature mechanism dominates the observed AHE, while the extrinsic contributions are much smaller. 
The intrinsic AHC increases as the film thickness increases, while the extrinsic AHC is thickness-independent. Such an enhanced intrinsic AHC in the MnPt$_3$ films is most likely attributed to the strain effect, implying that it serves as an effective method to tune the electronic band topology in the $X$Pt$_3$ topological semimetal.

\end{abstract}

\maketitle\enlargethispage{3pt}

\vspace{-5pt}
\section{\label{sec:Introduction}Introduction}\enlargethispage{8pt}

Anomalous Hall effect (AHE) has attracted great interest in recent years due to its intimate connection to the band topology of quantum materials~\cite{Nagaosa2010,Nakatsuji2015,Libor2022}. Topological magnetic materials usually host nontrivial band crossings (e.g., Weyl or Dirac points) near the Fermi level~\cite{Yan2017,Manna2018,Smejkal2018,Tokura2019,Wieder2021,Bernevig2022,He2022}.
The interplay between spin-orbit coupling (SOC) and magnetic order usually leads to large Berry curvatures and AHE in those materials, e.g., TbMn$_6$Sn$_6$~\cite{Yin2020,Wei2025}, Co$_3$Sn$_2$S$_2$~\cite{Liu2018a}, Mn$_3$$X$ ($X$ = Ge, Ir)~\cite{Nayak2016,MacDonald2014}, etc. 
In most of those magnetic materials, two different mechanisms account for the observed AHE. According to the phenomenological scaling between anomalous Hall conductivity (AHC) $\sigma_\mathrm{xy}^\mathrm{A}$ and electrical conductivity $\sigma_\mathrm{xx}$~\cite{Nagaosa2010,chen2021,yang2020,Wang2025}, 
the intrinsic Berry-curvature mechanism is dominant in the good-metal regime, where $\sigma_\mathrm{xy}^\mathrm{A}$ is almost a constant. On the contrary, the extrinsic skew-scattering- and side-jump mechanisms are at play in the high-conductivity- and bad-metal regimes, which lead to $\sigma_\mathrm{xy}^\mathrm{A}$ $\propto$ $\sigma_\mathrm{xx}$ and $\sigma_\mathrm{xy}^\mathrm{A}$ $\propto$ $\sigma_\mathrm{xx}^{1.6-1.8}$, respectively.

Binary Mn-Pt alloy films exhibit rich magnetic and transport properties, which play an important role in the field of spintronics~\cite{Chen2024}.  
For Pt concentration below 30\%, the Mn-rich alloys adopt a cubic Cu$_3$Au-type structure ($Pm$$\bar{3}$$m$, No.~221)~\cite{Kren1966}. In stoichiometric Mn$_3$Pt alloy, Pt and Mn atoms preferentially occupy the 1$a$ (0, 0, 0) and 3$c$ (0.5, 0.5, 0) Wyckoff positions, respectively. In this structure, Mn atoms form a kagome sublattice along the [111]-direction with a 120$^\circ$ spin configuration, also known as a non-collinear antiferromagnet~\cite{Long1991}. Despite its weak net magnetization, Mn$_3$Pt exhibits both AHE and spin Hall effect (SHE) even at room temperature, mostly arising from the Berry curvatures of the nontrivial electronic bands~\cite{Zhang2017,liu2018,Mukherjee2021,Zuniga2023,Chen2025,Xu2024,Zhao2022,Novakov2023}. The AHE in Mn$_3$Pt can be effectively tuned either by varying the alloy composition or by epitaxial strain~\cite{Sinha2025,An2020}.
In addition, Mn$_3$Pt has been extensively studied due to its strong SOC and symmetry-governed spin-current anisotropy~\cite{Cao2023,Qin2023}.

As the Pt concentration further increases from 30\% up to 60\%, Mn-Pt alloys adopt a tetragonal CuAu-type structure ($P4/mmm$, No.~123). For the stoichiometric MnPt alloy, Mn and Pt atoms prefer to occupy 1$a$ (0, 0, 0) and 1$d$ (0.5, 0.5, 0.5) Wyckoff positions, respectively.  
MnPt typically exhibits a collinear antiferromagnetic (AFM) structure~\cite{Kren1968}, and has been widely used in the spintronic devices. 
Due to its extremely high N\'eel temperature ($T_\mathrm{N} \approx $ 973\,K) and strong magnetocrystalline anisotropy~\cite{Kren1968}, MnPt is frequently employed as an AFM pinning layer in the spin-valve devices~\cite{Parkin2003,Iusipova2021}.

On the Pt-rich side, Mn-Pt alloys with Pt concentration ranging from 63 to 83\% restore to a cubic Cu$_3$Au-type structure. However, different from the Mn$_3$Pt case, the atomic positions of Mn and Pt are switched in  the  stoichiometric MnPt$_3$, namely, Mn and Pt occupy 3$c$ and 1$a$ Wyckoff positions, respectively. For both Cu$_3$Au- and CuAu-type %Mn-Pt films 
structures with different stoichiometry, site mixing has been frequently observed, which significantly affects their magnetic and transport properties~\cite{Sinha2025}. %~\cite{Ma2015,Lu2009}.  
While AFM Mn$_3$Pt and MnPt films have been widely explored, the Pt-rich MnPt$_3$ films remain largely unexplored. Different from Mn$_3$Pt and MnPt, bulk MnPt$_3$ shows a ferromagnetic (FM) ground state with a Curie temperature $T_\mathrm{C}$ $\sim$ 390\,K~\cite{Antonini1969}. Though a large magneto-optic Kerr rotation has been reported in MnPt$_3$ films~\cite{Wierman1997,Kato1995,Oppeneer1996}, their spin transport properties have not yet been reported to the best of our knowledge. Very recently, theoretical calculations have revealed that the $X$Pt$_3$ (X = V, Cr, and Mn) family is a topological semimetal, and exhibits anti-crossing gapped nodal lines near the Fermi level, which give rise to distinct Berry curvatures and thus AHE~\cite{Markou2021}. Among the three members, CrPt$_3$ has been experimentally verified to show a large AHC $\sigma_\mathrm{xy}^\mathrm{A} \sim$ 1750~$\mathrm{\Omega}^{-1}$cm$^{-1}$ that is comparable to the theoretical maximum value of 1965~$\mathrm{\Omega}^{-1}$cm$^{-1}$. However, its counterpart MnPt$_3$ remains largely unexplored.

Here, a series of MnPt$_3$ thin films with different thicknesses was epitaxially grown on the (001)-oriented MgO substrates. We report a systematic study of their magnetic and transport properties by means of magnetization, electrical resistivity, and Hall resistivity measurements. All the MnPt$_3$ films undergo a FM transition at $T_\mathrm{c}$ between 309 and 344 K. In the magnetically ordered state, distinct AHE was observed, which is predominantly governed by the intrinsic Berry curvature mechanism.

\section{Experimental details\label{sec:details}}\enlargethispage{8pt}

High-quality MnPt$_3$ films with varied thicknesses were epitaxially grown on the (001)-oriented MgO substrates by
magnetron co-sputtering of Mn and Pt metal targets in an ultrahigh vacuum chamber with a base pressure below 5 $\times$ 10$^{-8}$ Torr. Prior to the deposition, MgO substrates were annealed at 600\,$^\circ$C for 1 h to eliminate moisture and surface contaminants. 
Both Mn and Pt atoms were deposited in an argon atmosphere with a fixed pressure of 3 mTorr at 600\,$^\circ$C. To produce MnPt$_3$ (i.e., Mn$_{25}$Pt$_{75}$) films, the relative contents of Mn and Pt were controlled by adjusting the sputtering power of Mn and Pt targets 
following the equation 
\begin{equation}
	\label{eq:comp}
	v_\mathrm{Pt}=\frac{xM_\mathrm{Pt}\rho_\mathrm{Mn}}{(1-x)M_\mathrm{Mn}\rho_\mathrm{Pt}}v_\mathrm{Mn},
\end{equation}
where $v_\mathrm{Mn(Pt)}$, $M_\mathrm{Mn(Pt)}$, and $\rho_\mathrm{Mn(Pt)}$ denote the deposition rate, molar mass, and density of Mn and Pt targets, respectively. According to the above equation, the deposition rates of Mn and Pt targets were fixed to the values of  0.039\,nm s$^{-1}$ and 0.14\,nm s$^{-1}$, respectively. After the deposition, MnPt$_3$ films were annealed at the same temperature for an additional hour to improve their crystallinity. Finally, a 4-nm thick Pt capping layer was deposited at room temperature to protect the MnPt$_3$ films from oxidation.

The crystal structure and epitaxial nature of MnPt$_3$ films were characterized by Malvern Panalytical X’Pert high-resolution X-ray diffractometer (HRXRD) with Cu-K$_\alpha$ radiation ($\lambda$ = 1.5418\,\AA{}). Film thickness was determined by fitting the X-ray reflectivity (XRR) patterns. 
The magnetic properties of the MnPt$_3$ films were studied using a Quantum Design magnetic property measurement system. Measurements of
transverse Hall resistivity $\rho_\mathrm{yx}$ and longitudinal resistivity $\rho_\mathrm{xx}$ were carried out in a Quantum Design physical property measurement system. For the transport measurements, MnPt$_3$ films were patterned into a Hall-bar geometry (central area: 100\,{\textmu}m $\times$ 20\,{\textmu}m; electrodes: 20\,{\textmu}m $\times$ 20\,{\textmu}m) by the standard photolithography and Ar-ion-beam etching techniques.

%
%==== figure =============================%
\begin{figure}[!thp]
	\centering
	%\vspace{-1ex}%
	\includegraphics[width=0.48\textwidth,angle=0]{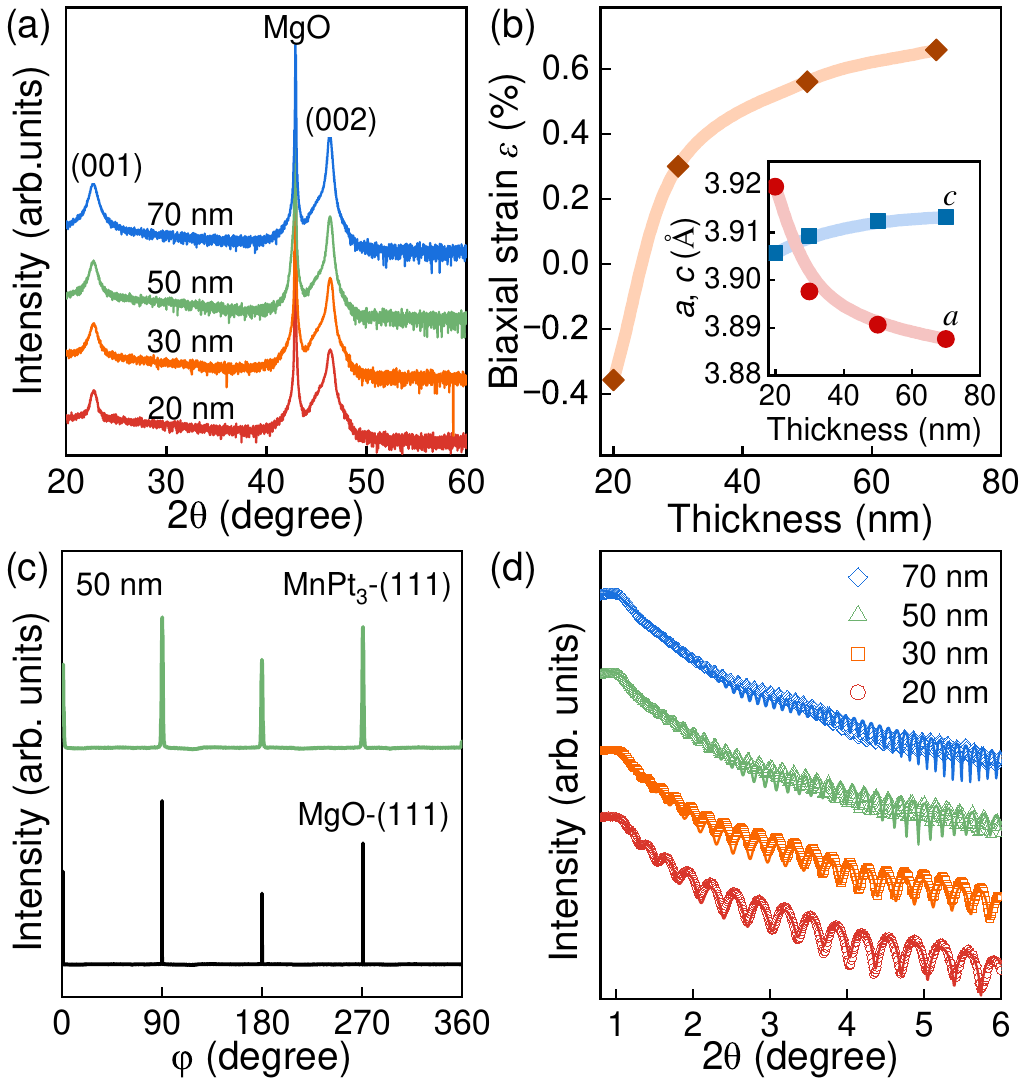}
	\caption{\label{fig:XRD}
		(a) XRD pattern of MnPt$_3$ films with different thicknesses. 
		(b) Biaxial strain as a function of thickness. The inset shows thickness-dependent in-plane and out-of-plane lattice parameters. 
		(c) $\varphi$-scan patterns for the 50-nm-thick MnPt$_3$ film and the MgO substrate. The analogous results for other MnPt$_3$ films are shown in Fig.~S3 in the Supplementary Materials~\cite{Supple}.
		(d) XRR patterns for MnPt$_3$ films with different thicknesses. Solid lines through the data represent fitting curves.
	}
\end{figure}
%=== end figure ==========================%
%

\section{Results and discussion\label{sec:results}}\enlargethispage{8pt}
%\subsection{Structural characterization\label{ssec:XRD}}

%==== figure =============================%
\begin{figure*}[!htp]
	\centering
	\vspace{-1ex}%
	\includegraphics[width=0.97\textwidth,angle=0]{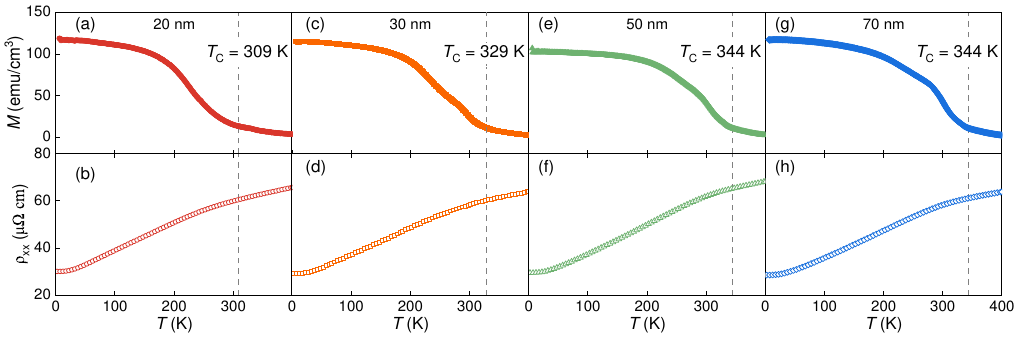}
	\caption{\label{fig:Tc} \tcr{
		Temperature-dependent magnetization $M(T)$ (a) and electrical resistivity $\rho_\mathrm{xx}(T)$ (b) for the 20-nm thick MnPt$_3$ film. The
		analogous results for the 30-, 50-, and 70-nm-thick MnPt$_3$ films are shown in panels (c)-(d), (e)-(f), and (g)-(h), respectively. 
		For $M(T)$ measurements, magnetization was collected by applying a field of $\mu_{0}H$ = 0.1\,T perpendicular to the film plane (i.e., $H$ $\parallel$ $c$) using field-cooled (FC) protocol. 
		When applying the magnetic field within the film plane (i.e., $H$ $\perp$ $c$), similar temperature-dependent $M(T)$ was observed for all the MnPt$_3$ films (see details in Fig.~S6 in the Supplementary Materials~\cite{Supple}).
		Magnetic transition temperatures $T_C$ determined from the derivatives of magnetization d$M$/d$T$ and electrical resistivity d$\rho_\mathrm{xx}$/d$T$ with respect to temperature (see Figs.~S7 and S8 in the Supplementary Materials~\cite{Supple}) are marked by the dashed lines. The contribution from the MgO substrate was subtracted from the measured $M(T)$ (see details in Fig.~S4 in the Supplementary Materials~\cite{Supple}).}
 	}
\end{figure*}
%=== end figure ==========================%

The crystal structure and the epitaxial nature of MnPt$_3$ films were characterized by HRXRD measurements. Figure~\ref{fig:XRD}(a) plots the HRXRD patterns for MnPt$_3$ films with different thicknesses. All the MnPt$_3$ films exhibit clear (001) and (002) reflections, 
consistent with a chemically ordered phase (see crystal structure in Fig.~S1 in the Supplementary Materials~\cite{Supple} and also references \cite{Fujishiro2021,Park2020} therein). 
The absence of foreign phases or misorientation suggests high crystalline quality of the deposited MnPt$_3$ films. The out-of-plane (i.e., $c$ axis) and the in-plane lattice parameters (i.e., $a$ axis) of MnPt$_3$ films were calculated according to the (001) and (022) reflections (see Fig.~S1 in the Supplementary Materials~\cite{Supple}), respectively. As shown in the inset of Fig.~\ref{fig:XRD}(b),
the $c$ axis increases as the film thickness increases, reaching 3.913~\AA{} for the 70-nm-thick MnPt$_3$ film. By contrast, the $a$ axis  shows an opposite thickness dependence and decreases to 3.888~\AA{} for the 70-nm-thick MnPt$_3$ film. Such thickness-dependent $a$ and $c$ axes imply an enhanced structural distortion and/or tetragonality in the MnPt$_3$ films. This is clearly reflected by the thickness-dependent biaxial strain $\varepsilon$, which is defined as $\varepsilon$ = $(c-a)/a$. As shown in Fig.~\ref{fig:XRD}b, the 20-nm thick MnPt$_3$ film shows a negative $\varepsilon$, consistent with the fact that $a$ axisis larger than $c$. As further increasing the thickness, $\varepsilon$ becomes positive and continually increases, reaching 0.64\% in the 70-nm-thick MnPt$_3$ film. The epitaxial nature of MnPt$_3$ films was characterized by $\varphi$-scan measurements with a 2$\theta$ value fixed at the (111) reflection of the MnPt$_3$ films and the MgO substrate.
The $\varphi$-scan patterns of the 50-nm-thick MnPt$_3$ are shown in Fig.~\ref{fig:XRD}(c), which confirm that MnPt$_3$ films were epitaxially grown on the MgO substrate with a cube-on-cube growth. Different from the MnPt$_3$ case, Fe-Rh and Mn-Rh alloy films were epitaxially grown on the MgO substrate with an in-plane 45$^\circ$ rotation~\cite{Zhu2023,Wang2025}. The epitaxial growth was furhter confirmed by reciprocal space mapping (RSM) measurements around the (113) reflection of both the MgO substrate and the MnPt$_3$ films (see Fig.~S2 in the Supplementary Materials~\cite{Supple}). The lattice parameters determined from RSM are consistent with those calculated from HRXRD patterns. The thickness of MnPt$_3$ films was determined by XRR measurements [see Fig.~\ref{fig:XRD}(d)]. The well defined finger oscillations indicate ideal flatness (with a roughness less than 1\,nm, see Table S1 in the Supplementary Materials~\cite{Supple}) and uniformity of the MnPt$_3$ films. 
\tcr{The estimated thicknesses of the MnPt$_3$ films are 21.0(1), 31.0(1), 51.7(3), and 73.4(2)\,nm, respectively}.
For simplicity, the thicknesses of 20, 30, 50, and 70\,nm are used in this paper.

Magnetic and transport properties of MnPt$_3$ films were first characterized by temperature-dependent magnetization $M(T)$ and electrical-resistivity $\rho_\mathrm{xx}(T)$ measurements. MnPt$_3$ films undergo a paramagnetic (PM) to FM transition at the Curie temperature $T_\mathrm{C}$ (see top panels in Fig.~\ref{fig:Tc}). The FM ground state is further supported by the distinct magnetic hysteresis in the field-dependent magnetization $M(H)$ (see below).
The electrical resistivity decreases upon cooling temperature, indicating metallic nature of MnPt$_3$ films (see bottom panels in Fig.~\ref{fig:Tc}). Resembling the $M(T)$ curves, $\rho_\mathrm{xx}(T)$ also exhibits a clear anomaly at $T_\mathrm{C}$. 
The FM transition temperatures $T_\mathrm{C}$ determined from the derivatives of magnetization d$M$/d$T$ and electrical resistivity d$\rho_\mathrm{xx}$/d$T$ with respect to temperature are highly consistent (see Figs.~S7 and S8 in the Supplementary Materials~\cite{Supple}). As marked by the dashed lines in Fig.~\ref{fig:Tc},  $T_\mathrm{C}$ increases from 309\,K to 344\,K as the thickness increases from 20\,nm to 70\,nm in the MnPt$_3$ films. It is noted that the Curie temperature of the 70-nm-thick MnPt$_3$ film is almost identical to the bulk value~\cite{Kurt1996}. 

\begin{figure*}[!htp]
	\centering
	\includegraphics[width=0.90\textwidth,angle= 0]{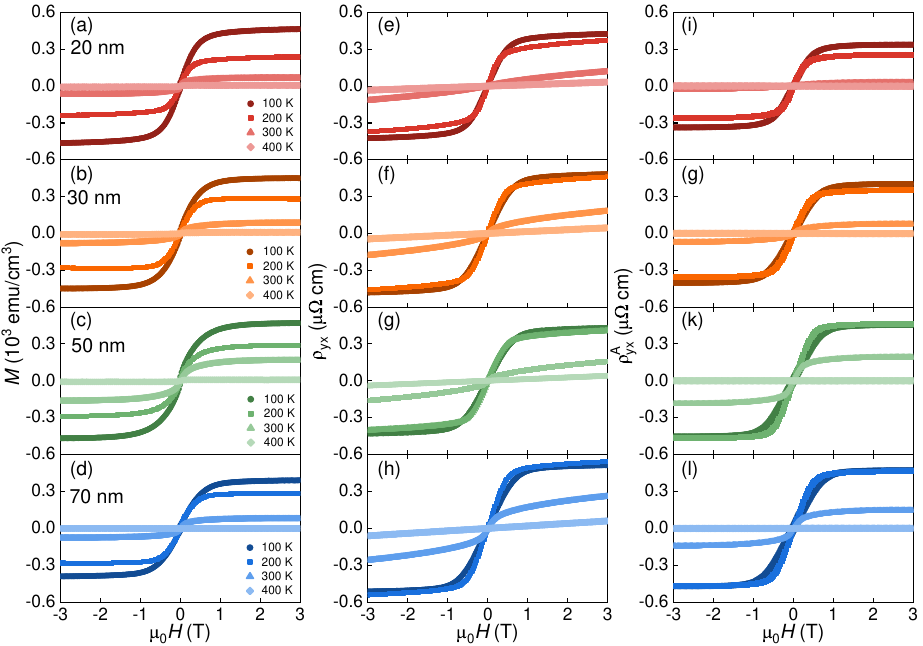}
	\caption{\label{fig:RH} \tcr{
		Field-dependent magnetization $M(H)$ collected at selected temperatures for MnPt$_3$ films with thicknesses of (a) 20, (b) 30, (c) 50, and (d) 70\,nm. The field-dependent Hall resistivity $\rho_\mathrm{yx}(H)$ and anomalous Hall resistivity $\rho^\mathrm{A}_\mathrm{yx}(H)$ are presented in panels (e)-(h) and (i)-(l), respectively. 
		Results at other temperatures are presented in Figs.~S10 and S11 in the Supplementary Materials~\cite{Supple}.
		For all those measurements, magnetic field was applied along the normal direction of the films, i.e., $H$ $\parallel$ $c$. The field-dependent in-plane magnetization is presented in Fig.~S6 in the Supplementary Materials~\cite{Supple}. The contribution from the MgO substrate was subtracted from the measured $M(H)$ (see details in Fig.~S5 in the Supplementary Materials~\cite{Supple})}.
	}
	
\end{figure*}

To further confirm the magnetic ground state of MnPt$_3$ films, field-dependent magnetization $M(H)$ up to 3\,T was measured at selected temperatures below 400\,K. As shown in Fig.~\ref{fig:RH}(a)-(d), $M(H)$ curves exhibit typical features for the FM thin films.  
For all the MnPt$_3$ films, both in-plane and out-of-plane magnetization data exhibit clear magnetic hysteresis loops near zero field (see Fig.~S9 in the Supplementary Materials~\cite{Supple}). In addition, the in-plane spontaneous magnetization is significantly larger than that of the out-of-plane,  
indicating an in-plane magnetic anisotropy of MnPt$_3$ films. The estimated saturation fields are {\textmu}$_0H_\mathrm{s}$ = 0.06 and 0.14\,T for $H \parallel c$ and $H \perp c$, respectively. 
In the PM state (e.g., $T$ = 400\,K), \tcr{the extremely weak spontaneous magnetization is most likely attributed to the remanent
magnetization at the interfaces, which was frequently observed in other Mn-based thin films~\cite{Ishino2018,Li2025}}.
Field-dependent Hall resistivity $\rho_\mathrm{yx}(H)$ of MnPt$_3$ films were measured at temperatures between 10 and 400\,K (see all the datasets in Figs.~S10 and S11 in the Supplementary Materials~\cite{Supple}). As shown in Fig.~\ref{fig:RH}(e)-(h), the $\rho_\mathrm{yx}(H)$ is linear in field in the PM state (e.g., 400\,K), where the ordinary Hall effect (OHE) is dominated for all the MnPt$_3$ films.  By contrast, as the temperature decreases below $T_\mathrm{C}$, $\rho_\mathrm{yx}(H)$ exhibits almost identical field responses as $M(H)$, implying that the AHE dominates the Hall resistivity. 

%\subsection{Magnetotransport properties\label{ssec:MH}}

To investigate the AHE of MnPt$_3$ films, Hall resistivity is analyzed following the equation $\rho_\mathrm{yx}(H) = \rho_\mathrm{yx}^\mathrm{O}(H) + \rho_\mathrm{yx}^\mathrm{A}(H)$, where $\rho_\mathrm{yx}^\mathrm{O}$ and $\rho_\mathrm{yx}^\mathrm{A}$ represent the ordinary and the anomalous Hall resistivity (AHR), respectively. For a single-band picture, the first OHE term $\rho_\mathrm{yx}^\mathrm{O}$ (= $R_0$$H$) is proportional to the applied magnetic field, and the second AHE term $\rho_\mathrm{yx}^\mathrm{A}$ (= $R_\mathrm{S}$$M$) is mostly determined by the magnetization. In real materials, $R_\mathrm{S}$ 
can be a constant or proportional to $\rho_\mathrm{xx}$ or $\rho_\mathrm{xx}^2$, depending on the dominant mechanism, e.g., intrinsic, side-jump, or skew scattering~\cite{Nagaosa2010}. The positive $R_0$ coefficients of $\rho_\mathrm{yx}^\mathrm{O}$ indicate the dominant \tcr{hole}-type carriers in all the MnPt$_3$ films. The carrier density $n$ calculated from the $R_0$ coefficient is summarized in Fig. S12 in the Supplementary Materials~\cite{Supple}. In the PM state, $n$ slightly decreases as temperature decreases. By contrast, $n$ shows an opposite temperature dependence in the FM state, which increases as the temperature decreases.
\tcr{In addition, the carrier density is largely enhanced as the film thickness increases in the FM state.} 
For example, $n$ increases from 0.91 $\times$ 10$^{29}$ m$^{-3}$ for the 20-nm-thick film to 1.65 $\times$ 10$^{29}$ m$^{-3}$ for the 70-nm-thick MnPt$_3$ film at $T$ = 10\,K. \tcr{In general, MnPt$_3$ films show larger carrier densities at temperatures well below $T_\mathrm{C}$. The band splitting caused by the FM order plays a key role in determining the band topology and density of states near the Fermi level, and thus, the carrier density in MnPt$_3$ films. Further theoretical work, including the band-structure calculations by considering the spin polarization, is highly desirable.}

\begin{figure}[!htp]
	\centering
	\includegraphics[width=0.40\textwidth,angle= 0]{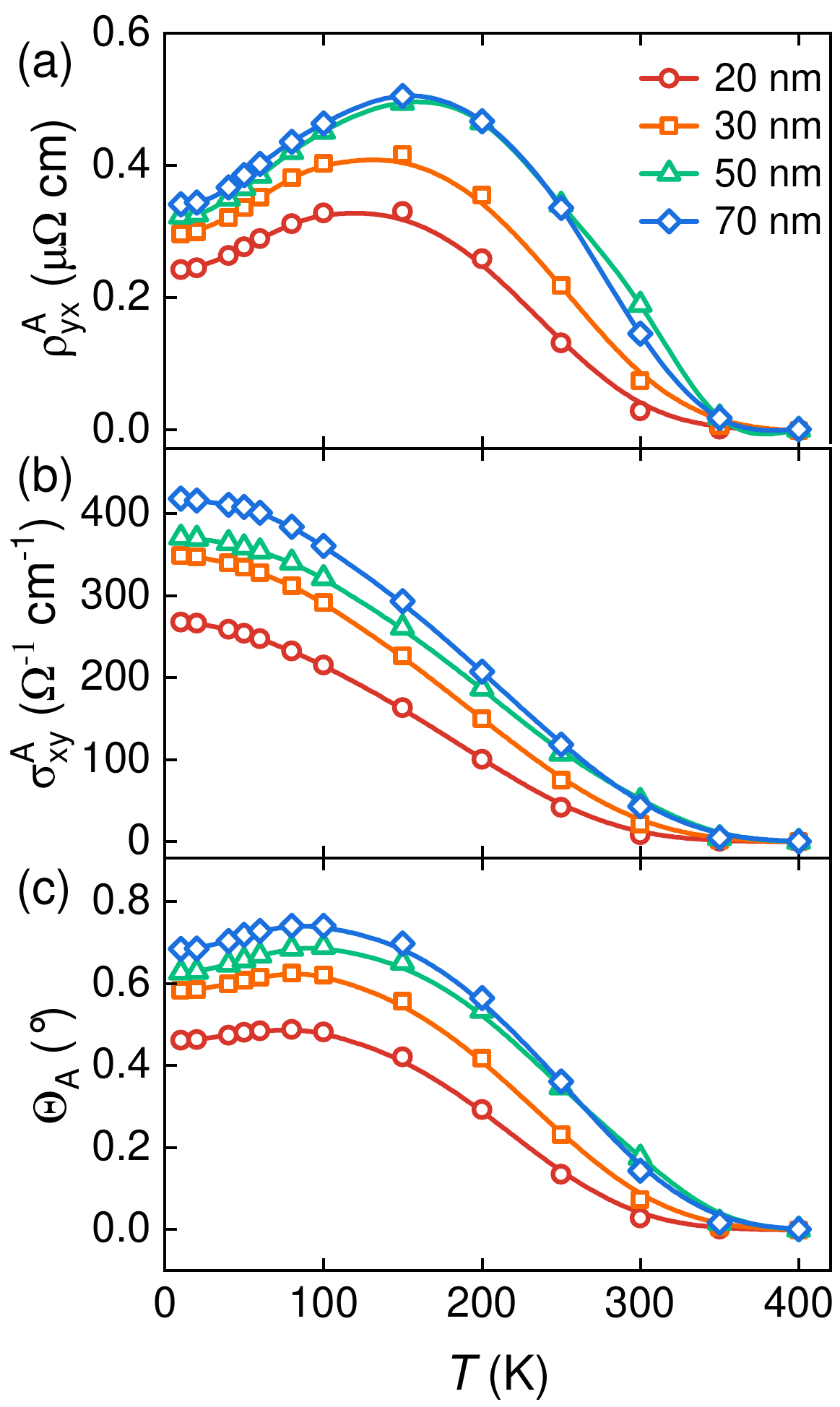}
	\caption{\label{fig:phase} \tcr{(a) Temperature-dependent anomalous Hall resistivity $\rho_\mathrm{yx}^\mathrm{A}$, (b) anomalous Hall conductivity $\sigma_\mathrm{xy}^\mathrm{A}$, and (c) anomalous Hall angle $\Theta_\mathrm{A}$ for MnPt$_3$ films with varied thicknesses.}}
\end{figure}

The AHR $\rho_\mathrm{yx}^\mathrm{A}$ of MnPt$_3$ films was extracted simply by subtracting the linear term. The resulting $\rho_\mathrm{yx}^\mathrm{A}(H)$ at some representative temperatures are shown in Fig.~\ref{fig:RH}(i)-(l). The $\rho_\mathrm{yx}^\mathrm{A}(H)$ also exhibits a clear magnetic hysteresis in the FM state (see Fig.~S13 in the Supplementary Materials~\cite{Supple}), resembling the $M(H)$ data in Fig.~\ref{fig:RH}(a)-(d). The derived AHR of MnPt$_3$ films with different thicknesses is summarized in Fig.~\ref{fig:phase}(a). For all the MnPt$_3$ films, $\rho_\mathrm{yx}^\mathrm{A}$ exhibits a non-monotonic temperature dependence, which increases as the temperature decreases until reaching a maximum value at $T$ = 150\,K, below which $\rho_\mathrm{yx}^\mathrm{A}$ starts to decrease again.
\tcr{Such temperature-dependent $\rho_\mathrm{yx}^\mathrm{A}(T)$ is attributed to the combined effect of saturation magnetization and electrical resistivity, which reflects the competition between the intrinsic and extrinsic mechanisms.}
For a given temperature, $\rho_\mathrm{yx}^\mathrm{A}$ increases as the film thickness increases. For example, $\rho_\mathrm{yx}^\mathrm{A}$ = 0.33, 0.42, 0.49, and 0.51~{\textmu}$\mathrm{\Omega}$ cm for the 20-, 30-, 50-, and 70-nm-thick MnPt$_3$ films at $T$ = 150\,K, respectively. The AHC $\sigma_\mathrm{xy}^\mathrm{A}$ was calculated according to $\sigma_\mathrm{xy}^\mathrm{A} = \rho_\mathrm{yx}^\mathrm{A}/[(\rho_\mathrm{yx}^\mathrm{A})^2 + (\rho_\mathrm{xx})^2]$, where $\rho_\mathrm{xx}$ is the zero-field electrical resistivity shown in Fig.~\ref{fig:Tc}. Different from $\rho_\mathrm{yx}^\mathrm{A}$, $\sigma_\mathrm{xy}^\mathrm{A}$ continuously increases as temperature decreases at $T < T_\mathrm{C}$ [Fig.~\ref{fig:phase}(b)], and starts to saturate at $T <$ 100\,K. 
\tcr{Such a temperature evolution of AHC is closely related to the temperature-dependent saturation magnetization $M_\mathrm{s}$ (see Fig.~S14 in the Supplementary Materials~\cite{Supple}) and electrical resistivity (Fig.~\ref{fig:Tc}), which determine the intrinsic and extrinsic AHC, respectively.}
The $\sigma_\mathrm{xy}^\mathrm{A}$ reaches 268, 349, 371, and 419 $\mathrm{\Omega}^{-1}$cm$^{-1}$ at $T$ = 10\,K for the 20-, 30-, 50-, and 70-nm-thick MnPt$_3$ films, respectively. 
The anomalous Hall angle $\Theta_\mathrm{A}$ [$\equiv \tan^{-1}(\sigma_\mathrm{xy}^\mathrm{A}/\sigma_\mathrm{xx})$]
shows similar temperature dependence as $\sigma_\mathrm{xy}^\mathrm{A}$ [Fig.~\ref{fig:phase}(c)]. Resembling the  $\rho_\mathrm{yx}^\mathrm{A}$, both $\sigma_\mathrm{xy}^\mathrm{A}$ and $\Theta_\mathrm{A}$ increase as the film thickness increases. 
The maximum of $\Theta_\mathrm{A}$ = 0.74$^\circ$ was observed in the 70-nm-thick MnPt$_3$ film.
At room temperature, the $\Theta_\mathrm{A}$ $\approx$ 0.14$^\circ$ of the MnPt$_3$ films is comparable to the CrPt$_3$ films~\cite{Markou2021}, but is much larger than other Mn-based metallic thin films. For example, the non-collinear antiferromagnet Mn$_3$Ir exhibits a much smaller $\Theta_\mathrm{A}$ = 0.03$^\circ$~\cite{Kobayashi2022}.

%

%\subsection{Magnetoresistivity and Hall resistivity\label{ssec:Electrical}}

%
%
\begin{figure}[!htp]
	\centering
	\includegraphics[width=0.49\textwidth,angle= 0]{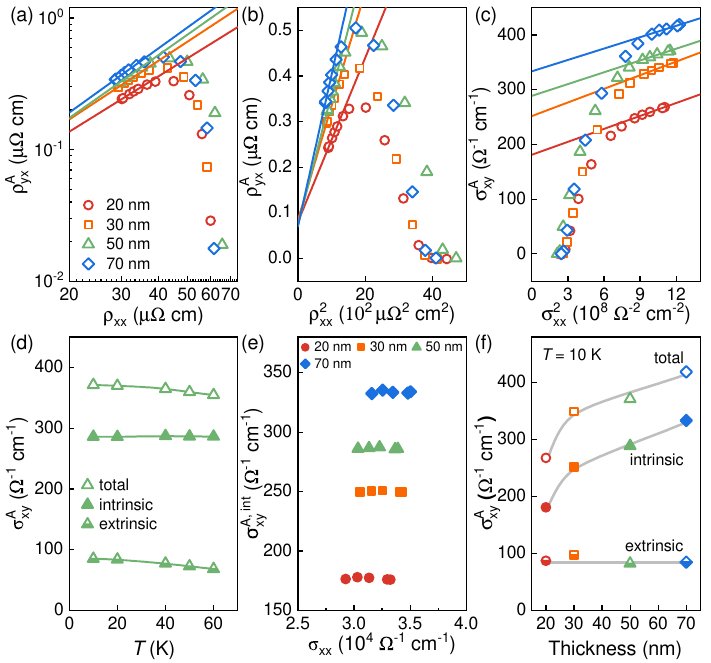}
	\caption{\label{fig:AHC}  
		(a) Anomalous Hall resistivity $\rho^\mathrm{A}_\mathrm{yx}$ versus electrical resistivity $\rho_\mathrm{xx}$ on logarithmic scales for MnPt$_3$ films with varied thicknesses. Solid lines represent power-law fits (i.e., $\rho^\mathrm{A}_\mathrm{xy}$ $\propto$  $\rho^\alpha_\mathrm{xx}$). (b) $\rho^\mathrm{A}_\mathrm{yx}$ versus $\rho^2_\mathrm{xx}$ for MnPt$_3$ films with varied thicknesses. (c) Anomalous Hall conductivity $\sigma^\mathrm{A}_\mathrm{xy}$ versus $\sigma^2_\mathrm{xx}$. Solid lines in panels (b) and (c) represent linear fits. (d) The total $\sigma_\mathrm{xy}^\mathrm{A,tot.}$, and the extracted intrinsic $\sigma_\mathrm{xy}^\mathrm{A,int.}$ and extrinsic AHC $\sigma_\mathrm{xy}^\mathrm{A,ext.}$ versus temperature below 60\,K for the 50-nm-thick film. Similar features were observed for other MnPt$_3$ films (see Fig.~S15 in the Supplementary Materials~\cite{Supple}).
		(e) Intrinsic AHC $\sigma_\mathrm{xy}^\mathrm{A,int.}$ versus  $\sigma_\mathrm{xx}$ for MnPt$_3$ films with varied thicknesses.  (f) Thickness-dependent intrinsic and extrinsic AHC for MnPt$_3$ films at $T$ = 10\,K. The AHC at other temperatures exhibits similar behaviors. 
	}
\end{figure}

Now we discuss the origins of AHE in the epitaxial MnPt$_3$ films. In general, the AHE is attributed to both intrinsic (i.e., Berry curvature) and extrinsic (skew-scattering or side-jump) mechanisms~\cite{Nagaosa2010}. 
Figure~\ref{fig:AHC}(a) plots the $\rho_\mathrm{yx}^\mathrm{A}$ versus the $\rho_\mathrm{xx}$ for MnPt$_3$ films with different thicknesses. 
As the temperature approaches 150\,K, where the magnetization starts to decrease rapidly (see Fig.~\ref{fig:Tc} and Fig.~S6 in the Supplementary Materials~\cite{Supple}), $\rho_\mathrm{yx}^\mathrm{A}$ exhibits a distinct drop for all the MnPt$_3$ films. By contrast, a power law scale (i.e., $\rho^\mathrm{A}_\mathrm{xy}$ $\propto$  $\rho^\alpha_\mathrm{xx}$) can be clearly identified at $T <$ 100\,K [see solid lines in Fig.~\ref{fig:AHC}(a)]. The derived $\alpha$ values are listed in Table~\ref{tab:parameter}, which slightly increases from 1.40 to 1.62 as the thickness increases. For all the MnPt$_3$ films, $\alpha$ value is clearly between 1 and 2, implying both intrinsic and extrinsic contributions to the observed AHE~\cite{Nagaosa2010}.
The extrinsic side-jump contribution to the AHC should be the magnitude of
$\frac{e^2}{ha}$$\frac{\varepsilon_{\mathrm{SO}}}{E_{\mathrm{F}}}$
for three-dimensional systems
~\cite{Lewiner1973,Onoda2006}, where $e$ is the electron charge, $h$ the Planck constant, $a$ the lattice constant, $\varepsilon_{\mathrm{SO}}$ the spin–orbit interaction energy, and $E_{\mathrm{F}}$ the Fermi energy. 
Since the $\frac{\varepsilon_{\mathrm{SO}}}{E_{\mathrm{F}}}$ is usually less than 0.01 for metallic ferromagnets~\cite{Wang2016,Kim2018,Roy2020,Sunil2025}, the estimated extrinsic side-jump contribution is on the order of 10~$\mathrm{\Omega}^{-1}$cm$^{-1}$ for MnPt$_3$ films, which is negligible compared with the measured AHC in Fig.~\ref{fig:phase}(b).

As an alternative, the Tian-Ye-Jin (TYJ) model was applied to distinguish the intrinsic contribution from the extrinsic contribution to the AHE in MnPt$_3$ films~\cite{Tian2009}.
Figure~\ref{fig:AHC}(b) plots the $\rho_\mathrm{yx}^\mathrm{A}$ versus $\rho_\mathrm{xx}^2$ for MnPt$_3$ films with different thicknesses. The solid lines are fits to the equation
\begin{equation}
	\label{eq:TYJ}
 \rho_\mathrm{yx}^\mathrm{A} = a_1\rho_\mathrm{xx,0} + b_1 \rho_\mathrm{xx}^2,
\end{equation}
where $\rho_\mathrm{xx,0}$ is the residual resistivity at zero temperature, $a_1$ is the skew scattering coefficient, and $b_1$ ($\equiv \sigma_\mathrm{xy}^\mathrm{A,int.}$) is the intrinsic AHC. 
\tcr{
As the temperature increases above 60\,K,  $\rho_\mathrm{yx}^\mathrm{A}$ undergoes a sudden drop, preventing the TYJ analysis. As a consequence, the TYJ model is limited to $T \le$ 60\,K for MnPt$_3$ films. This is most likely due to the significant change in the saturation magnetization $M_\mathrm{s}$ in this temperature range (see Fig.~S14 in the Supplementary Materials~\cite{Supple}).}
The derived $a_1$ and $b_1$ pkarameters are listed in Table~\ref{tab:parameter} for all the MnPt$_3$ films. Both the $a_1$ coefficient and the $\rho_\mathrm{xx,0}$ are almost independent of film thickness. By contrast, the $b_1$ parameter continuously increases from 180 to 333~$\mathrm{\Omega}^{-1}$cm$^{-1}$ as the film thickness increases from 20\,nm to 70\,nm. These results suggest that while the intrinsic contribution to the AHE is enhanced by increasing the film thickness, the extrinsic contribution is almost unchanged.

%==== Table =============================%
\begin{table}[!th]
	\centering
	\caption{\label{tab:parameter} Summary of the fitting parameters using TYJ model for MnPt$_3$ films with varied thicknesses. In is noted that $b_1$ and $d_1$  parameters are identical to the intrinsic AHC $\sigma_\mathrm{xy}^\mathrm{A,int.}$.} 
	\begin{ruledtabular}
		\begin{tabular}{lcccc}
			Thickness (nm)      &    20          &   30         &   50       &    70      \\   \hline  
			$\alpha$                 &    1.40(2)     &   1.48(1)    &   1.59(2)  & 1.62(3)    \\ %\hline 
			$a_1$ ($10^{-3}$)     &    2.65(14)    & 2.87(12)     &   2.46(12) & 2.45(24)   \\ 
			$\rho_\mathrm{xx,0}$ ({\textmu}$\mathrm{\Omega}$ cm) &   29.95(1)      & 29.04(1)       & 29.40(1)      &  28.47(1)    \\ 
			$b_1$ ($\mathrm{\Omega}^{-1}$cm$^{-1}$)  & 180(4)    &   251(4)   & 288(4) & 333(8)     \\ 
			$c_1$ ($10^{-3}$)                        & 2.8(1)    &   2.9(1)   & 2.5(1) & 2.4(2)       \\ 
			$d_1$ ($\mathrm{\Omega}^{-1}$cm$^{-1}$)  & 181(4)    &   252(4)   & 289(4) & 334(7)     \\ 
		\end{tabular}	
	\end{ruledtabular}
\end{table}
%=== end table ==========================%

The TYJ model also can  be expressed in a conductivity form
%%%%%%%%%%%%%%%%%
\begin{equation}
  \sigma_\mathrm{xy}^\mathrm{A} = -c_1\sigma_\mathrm{xx,0}^{-1}\sigma_\mathrm{xx}^{2} - d_1,
\end{equation}
%%%%%%%%%%%%%%%
where $\sigma_\mathrm{xx,0} = 1/\rho_\mathrm{xx,0}$ is the residual conductivity~\cite{Shen2020}. The first and the second term denotes the extrinsic ($\sigma_\mathrm{xy}^\mathrm{A,ext.}$) and intrinsic AHC ($\sigma_\mathrm{xy}^\mathrm{A,int.}$), respectively. Figure~\ref{fig:AHC}(c) plots the $\sigma_\mathrm{xy}^\mathrm{A}$ versus $\sigma_\mathrm{xx}^2$ for MnPt$_3$ films with different thicknesses. Resembling the $\rho_\mathrm{yx}^\mathrm{A}$ in Fig.~\ref{fig:AHC}(b), $\sigma_\mathrm{xy}^\mathrm{A}$ also shows a linear $\sigma_\mathrm{xx}^2$ dependence at temperatures below 60\,K, as shown by solid lines in Fig.~\ref{fig:AHC}(c). The derived $c_1$ and $d_1$ parameters are summarized in Table~\ref{tab:parameter}. Again, $\sigma_\mathrm{xy}^\mathrm{A,ext.}$ is almost independent of thickness, while $\sigma_\mathrm{xy}^\mathrm{A,int.}$ ($\equiv$ $d_1$) increases as the film thickness increases, being identical to the $b_1$ parameter in Fig.~\ref{fig:AHC}(b). According to the parameters listed in Table~\ref{tab:parameter}, $\sigma_\mathrm{xy}^\mathrm{A,ext.}$ was calculated at different temperatures below 60\,K. The temperature-dependent $\sigma_\mathrm{xy}^\mathrm{A,int.}$ was extracted by subtracting the measured AHC $\sigma_\mathrm{xy}^\mathrm{A,tot.}$ from the calculated $\sigma_\mathrm{xy}^\mathrm{A,ext.}$.
For all the MnPt$_3$ films [see Fig.~\ref{fig:AHC}(d) and Fig.~S15 in the Supplementary Materials~\cite{Supple}],
intrinsic AHC $\sigma_\mathrm{xy}^\mathrm{A,int.}$ is almost independent of temperature, while extrinsic AHC $\sigma_\mathrm{xy}^\mathrm{A,ext.}$ decreases slightly with increasing temperature. 

The extracted $\sigma_\mathrm{xy}^\mathrm{A,int.}$ for all the MnPt$_3$ films at different temperatures is summarized in Fig.~\ref{fig:AHC}(e). As expected from the TYJ model, $\sigma_\mathrm{xy}^\mathrm{A,int.}$ of MnPt$_3$ films with varied thicknesses is independent of $\sigma_\mathrm{xx}$. Figure~\ref{fig:AHC}(f) plots the $\sigma_\mathrm{xy}^\mathrm{A,tot.}$, $\sigma_\mathrm{xy}^\mathrm{A,ext.}$, and $\sigma_\mathrm{xy}^\mathrm{A,int.}$ as a function film thickness at $T$ = 10\,K. Clearly,  $\sigma_\mathrm{xy}^\mathrm{A,ext.}$ is independent of film thickness, while $\sigma_\mathrm{xy}^\mathrm{A,int.}$ is largely enhanced by increasing the film thickness, reaching 334 $\mathrm{\Omega}^{-1}$cm$^{-1}$ in the 70-nm thick MnPt$_3$ film,
which is significantly larger than the non-collinear AFM Mn$_3$Pt films, the latter show an AHC of $\sim$98 $\mathrm{\Omega}^{-1}$cm$^{-1}$~\cite{liu2018}. \tcr{Moreover, $\sigma_\mathrm{xy}^\mathrm{A,int.}$ is much larger than $\sigma_\mathrm{xy}^\mathrm{A,ext.}$ for all the MnPt$_3$ films.
The intrinsic contribution increases from 68\% to 80\% of the total AHC as the film thickness increases from 20 to 70\,nm.}
Therefore, the intrinsic Berry-curvature mechanism is dominant for the observed AHE in MnPt$_3$ epitaxial films, which is further supported by the $\sigma_\mathrm{xy}$ vs. $\sigma_\mathrm{xx}$ phase diagram (see Fig.~S16 in the Supplementary Materials~\cite{Supple}).
Since both magnetization and magnetic anisotropy of MnPt$_3$ films are not expected to change significantly with increasing film thickness (see Fig.~\ref{fig:RH}, Figs.~S6 and S9 in the Supplementary Materials~\cite{Supple}), the enhanced intrinsic AHE is most likely attributed to the evolution of electronic band structures and the associated Berry curvatures near the Fermi energy  $E_\mathrm{F}$, both of which can be tuned by the strain effect [see Fig.~\ref{fig:XRD}(b)]. \tcr{Such a conclusion is further supported by the plot of $\sigma_\mathrm{xy}^\mathrm{A,int.}$ versus the biaxial strain $\varepsilon$ in Fig.~S17 in the Supplementary Materials~\cite{Supple}, which shows a strong linear correlation. It could be interesting to grow MnPt$_3$ film on substrates with different lattice mismatches and to investigate their intrinsic AHE.}
The thickness-dependent $\sigma_\mathrm{xy}^\mathrm{A,int.}$ might also be affected by the chemical ordering.  
As shown in Fig.~S17 in the Supplementary Materials~\cite{Supple}, the chemical ordering parameter $S$ increases as the film thickness. 
\tcr{Similar results have been observed in the FePt, Mn$_{1.5}$Ga, Co$_2$FeSi$_{0.6}$Al$_{0.4}$, and Co$_2$FeGa$_{0.5}$Ge$_{0.5}$
films~\cite{Chen2011,Zhu2014,Zhu2012,Zhu2013,Vidal2011,He2012}, where the influence of chemical ordering on the intrinsic AHE is attributed to the modification of Fermi-surface topology~\cite{Chen2011,Vidal2011}. Besides, the chemical ordering may also affect the local SOC and  crystal symmetries, both of which play important roles in determining the intrinsic AHC in magnetic materials.}
In addition to the strain effect, the chemical ordering might shift the Fermi energy as well.  
According to the theoretical calculations~\cite{Markou2021}, the Berry‑curvature‑driven AHC reaches a maximum value of $\sim$2000 $\mathrm{\Omega}^{-1}$cm$^{-1}$ at $\sim$200\,meV below the Fermi energy of MnPt$_3$. Therefore, the shift of Fermi enegry can effectively tune the Berry curvature and thus the intrinsic AHC in MnPt$_3$. Indeed, the $E_\mathrm{F}$ should be shifted as the film thickness increases, reflected by a largely enhanced carrier density at low temperatures for the thick MnPt$_3$ films (see Fig.~S12 in the Supplementary Materials~\cite{Supple}). The shift of $E_\mathrm{F}$ could be attributed to the combined effects of strain, structure distortion, and chemical ordering. Further theoretical work, including the band-structure calculations by considering the carrier doping or structure distortion, is highly desirable.

\section{\label{ssec:Sum}CONCLUSION}\enlargethispage{8pt}
To summarize, a series of epitaxial MnPt$_3$ films with different thicknesses was grown on the (001)-oriented MgO substrates. By performing systematic magnetization and transport measurements, the thickness evolution of magnetic transition temperatures and anomalous transport properties of MnPt$_3$ films was established. All these anomalous transport properties, including resistivity, conductivity, and angle, are strongly correlated with the magnetic properties of MnPt$_3$ films. In the ferromagnetically ordered state, MnPt$_3$ films exhibit a distinct AHE. The scaling analysis using the TYJ model suggests that the intrinsic Berry-curvature mechanism is dominant in the observed AHE, while the extrinsic contributions are much smaller. The intrinsic AHC increases as the film thickness increases, while the extrinsic AHC is independent of the film thickness. Such an enhanced intrinsic AHC in the MnPt$_3$ films is most likely attributed to the strain effect, implying that strain engineering represents one of the effective methods to tune the electronic band topology in the topological magnetic semimetals.

\vspace{1pt}
\begin{acknowledgments}
We thank Dr. Jinying Yang for the fruitful discussions.
We acknowledge the support from Shanghai Synchrotron Radiation Facility (SSRF). Some of the transport data were collected on the a Quantum Design physical property measurement system at the RIXS station in SSRF.
This work was supported by the National Natural Science Foundation of China (Grant Nos. 12374105, 12274125, 12474126, 12350710785, and 12561160109), 
 and the Fundamental Research Funds for the Central Universities. J.M. acknowledges the financial support from the National Natural Science Foundation of China (Grant Nos. U2032213 and 12334008). B.L. acknowledges support from the National Natural Science Foundation of China (Grant Nos. 12374063 and 92565305), the Ministry of Science and Technology of China (Grant No. 2023YFA1407400), the Shanghai Natural Science Fund for Original Exploration Program (Grant No. 23ZR1479900), and the Cultivation Project of Shanghai Research Center for Quantum Sciences (Grant No. ZPY2024).
\end{acknowledgments}

%\appendix
%\section{\label{appendix} Frequency detuning of the NMR resonant circuit}

%\begin{footnotesize}
\bibliography{MnPt3.bib}

%apsrev4-2.bst 2019-01-14 (MD) hand-edited version of apsrev4-1.bst
%Control: key (0)
%Control: author (8) initials jnrlst
%Control: editor formatted (1) identically to author
%Control: production of article title (0) allowed
%Control: page (0) single
%Control: year (1) truncated
%Control: production of eprint (0) enabled
\begin{thebibliography}{63}%
\makeatletter
\providecommand \@ifxundefined [1]{%
 \@ifx{#1\undefined}
}%
\providecommand \@ifnum [1]{%
 \ifnum #1\expandafter \@firstoftwo
 \else \expandafter \@secondoftwo
 \fi
}%
\providecommand \@ifx [1]{%
 \ifx #1\expandafter \@firstoftwo
 \else \expandafter \@secondoftwo
 \fi
}%
\providecommand \natexlab [1]{#1}%
\providecommand \enquote  [1]{``#1''}%
\providecommand \bibnamefont  [1]{#1}%
\providecommand \bibfnamefont [1]{#1}%
\providecommand \citenamefont [1]{#1}%
\providecommand \href@noop [0]{\@secondoftwo}%
\providecommand \href [0]{\begingroup \@sanitize@url \@href}%
\providecommand \@href[1]{\@@startlink{#1}\@@href}%
\providecommand \@@href[1]{\endgroup#1\@@endlink}%
\providecommand \@sanitize@url [0]{\catcode `\\12\catcode `\$12\catcode
  `\&12\catcode `\#12\catcode `\^12\catcode `\_12\catcode `\%12\relax}%
\providecommand \@@startlink[1]{}%
\providecommand \@@endlink[0]{}%
\providecommand \url  [0]{\begingroup\@sanitize@url \@url }%
\providecommand \@url [1]{\endgroup\@href {#1}{\urlprefix }}%
\providecommand \urlprefix  [0]{URL }%
\providecommand \Eprint [0]{\href }%
\providecommand \doibase [0]{https://doi.org/}%
\providecommand \selectlanguage [0]{\@gobble}%
\providecommand \bibinfo  [0]{\@secondoftwo}%
\providecommand \bibfield  [0]{\@secondoftwo}%
\providecommand \translation [1]{[#1]}%
\providecommand \BibitemOpen [0]{}%
\providecommand \bibitemStop [0]{}%
\providecommand \bibitemNoStop [0]{.\EOS\space}%
\providecommand \EOS [0]{\spacefactor3000\relax}%
\providecommand \BibitemShut  [1]{\csname bibitem#1\endcsname}%
\let\auto@bib@innerbib\@empty
%</preamble>
\bibitem [{\citenamefont {Nagaosa}\ \emph {et~al.}(2010)\citenamefont
  {Nagaosa}, \citenamefont {Sinova}, \citenamefont {Onoda}, \citenamefont
  {MacDonald},\ and\ \citenamefont {Ong}}]{Nagaosa2010}%
  \BibitemOpen
  \bibfield  {author} {\bibinfo {author} {\bibfnamefont {N.}~\bibnamefont
  {Nagaosa}}, \bibinfo {author} {\bibfnamefont {J.}~\bibnamefont {Sinova}},
  \bibinfo {author} {\bibfnamefont {S.}~\bibnamefont {Onoda}}, \bibinfo
  {author} {\bibfnamefont {A.~H.}\ \bibnamefont {MacDonald}},\ and\ \bibinfo
  {author} {\bibfnamefont {N.~P.}\ \bibnamefont {Ong}},\ }\bibfield  {title}
  {\bibinfo {title} {Anomalous {Hall} effect},\ }\href
  {https://doi.org/10.1103/RevModPhys.82.1539} {\bibfield  {journal} {\bibinfo
  {journal} {Rev. Mod. Phys.}\ }\textbf {\bibinfo {volume} {82}},\ \bibinfo
  {pages} {1539} (\bibinfo {year} {2010})}\BibitemShut {NoStop}%
\bibitem [{\citenamefont {Nakatsuji}\ \emph {et~al.}(2015)\citenamefont
  {Nakatsuji}, \citenamefont {Kiyohara},\ and\ \citenamefont
  {Higo}}]{Nakatsuji2015}%
  \BibitemOpen
  \bibfield  {author} {\bibinfo {author} {\bibfnamefont {S.}~\bibnamefont
  {Nakatsuji}}, \bibinfo {author} {\bibfnamefont {N.}~\bibnamefont
  {Kiyohara}},\ and\ \bibinfo {author} {\bibfnamefont {T.}~\bibnamefont
  {Higo}},\ }\bibfield  {title} {\bibinfo {title} {Large anomalous {Hall}
  effect in a non-collinear antiferromagnet at room temperature},\ }\href
  {https://doi.org/10.1038/nature15723} {\bibfield  {journal} {\bibinfo
  {journal} {Nature}\ }\textbf {\bibinfo {volume} {527}},\ \bibinfo {pages}
  {212} (\bibinfo {year} {2015})}\BibitemShut {NoStop}%
\bibitem [{\citenamefont {Šmejkal}\ \emph {et~al.}(2022)\citenamefont
  {Šmejkal}, \citenamefont {MacDonald}, \citenamefont {Sinova}, \citenamefont
  {Nakatsuji},\ and\ \citenamefont {Jungwirth}}]{Libor2022}%
  \BibitemOpen
  \bibfield  {author} {\bibinfo {author} {\bibfnamefont {L.}~\bibnamefont
  {Šmejkal}}, \bibinfo {author} {\bibfnamefont {A.~H.}\ \bibnamefont
  {MacDonald}}, \bibinfo {author} {\bibfnamefont {J.}~\bibnamefont {Sinova}},
  \bibinfo {author} {\bibfnamefont {S.}~\bibnamefont {Nakatsuji}},\ and\
  \bibinfo {author} {\bibfnamefont {T.}~\bibnamefont {Jungwirth}},\ }\bibfield
  {title} {\bibinfo {title} {Anomalous {Hall} antiferromagnets},\ }\href
  {https://doi.org/10.1038/s41578-022-00430-3} {\bibfield  {journal} {\bibinfo
  {journal} {Nat. Rev. Mater.}\ }\textbf {\bibinfo {volume} {7}},\ \bibinfo
  {pages} {482} (\bibinfo {year} {2022})}\BibitemShut {NoStop}%
\bibitem [{\citenamefont {Yan}\ and\ \citenamefont {Felser}(2017)}]{Yan2017}%
  \BibitemOpen
  \bibfield  {author} {\bibinfo {author} {\bibfnamefont {B.}~\bibnamefont
  {Yan}}\ and\ \bibinfo {author} {\bibfnamefont {C.}~\bibnamefont {Felser}},\
  }\bibfield  {title} {\bibinfo {title} {Topological materials: {Weyl}
  semimetals},\ }\href
  {https://doi.org/10.1146/annurev-conmatphys-031016-025458} {\bibfield
  {journal} {\bibinfo  {journal} {Annu. Rev. Condens. Matter Phys.}\ }\textbf
  {\bibinfo {volume} {8}},\ \bibinfo {pages} {337} (\bibinfo {year}
  {2017})}\BibitemShut {NoStop}%
\bibitem [{\citenamefont {Manna}\ \emph {et~al.}(2018)\citenamefont {Manna},
  \citenamefont {Sun}, \citenamefont {Muechler}, \citenamefont {K\"{u}bler},\
  and\ \citenamefont {Felser}}]{Manna2018}%
  \BibitemOpen
  \bibfield  {author} {\bibinfo {author} {\bibfnamefont {K.}~\bibnamefont
  {Manna}}, \bibinfo {author} {\bibfnamefont {Y.}~\bibnamefont {Sun}}, \bibinfo
  {author} {\bibfnamefont {L.}~\bibnamefont {Muechler}}, \bibinfo {author}
  {\bibfnamefont {J.}~\bibnamefont {K\"{u}bler}},\ and\ \bibinfo {author}
  {\bibfnamefont {C.}~\bibnamefont {Felser}},\ }\bibfield  {title} {\bibinfo
  {title} {{Heusler}, {Weyl} and {Berry}},\ }\href
  {https://doi.org/10.1038/s41578-018-0036-5} {\bibfield  {journal} {\bibinfo
  {journal} {Nat. Rev. Mater.}\ }\textbf {\bibinfo {volume} {3}},\ \bibinfo
  {pages} {244} (\bibinfo {year} {2018})}\BibitemShut {NoStop}%
\bibitem [{\citenamefont {\v{S}mejkal}\ \emph {et~al.}(2018)\citenamefont
  {\v{S}mejkal}, \citenamefont {Mokrousov}, \citenamefont {Yan},\ and\
  \citenamefont {MacDonald}}]{Smejkal2018}%
  \BibitemOpen
  \bibfield  {author} {\bibinfo {author} {\bibfnamefont {L.}~\bibnamefont
  {\v{S}mejkal}}, \bibinfo {author} {\bibfnamefont {Y.}~\bibnamefont
  {Mokrousov}}, \bibinfo {author} {\bibfnamefont {B.}~\bibnamefont {Yan}},\
  and\ \bibinfo {author} {\bibfnamefont {A.~H.}\ \bibnamefont {MacDonald}},\
  }\bibfield  {title} {\bibinfo {title} {Topological antiferromagnetic
  spintronics},\ }\href {https://doi.org/10.1038/s41567-018-0064-5} {\bibfield
  {journal} {\bibinfo  {journal} {Nat. Phys.}\ }\textbf {\bibinfo {volume}
  {14}},\ \bibinfo {pages} {242} (\bibinfo {year} {2018})}\BibitemShut
  {NoStop}%
\bibitem [{\citenamefont {Tokura}\ \emph {et~al.}(2019)\citenamefont {Tokura},
  \citenamefont {Yasuda},\ and\ \citenamefont {Tsukazaki}}]{Tokura2019}%
  \BibitemOpen
  \bibfield  {author} {\bibinfo {author} {\bibfnamefont {Y.}~\bibnamefont
  {Tokura}}, \bibinfo {author} {\bibfnamefont {K.}~\bibnamefont {Yasuda}},\
  and\ \bibinfo {author} {\bibfnamefont {A.}~\bibnamefont {Tsukazaki}},\
  }\bibfield  {title} {\bibinfo {title} {Magnetic topological insulators},\
  }\href {https://doi.org/10.1038/s42254-018-0011-5} {\bibfield  {journal}
  {\bibinfo  {journal} {Nat. Rev. Phys.}\ }\textbf {\bibinfo {volume} {1}},\
  \bibinfo {pages} {126} (\bibinfo {year} {2019})}\BibitemShut {NoStop}%
\bibitem [{\citenamefont {Wieder}\ \emph {et~al.}(2021)\citenamefont {Wieder},
  \citenamefont {Bradlyn}, \citenamefont {Cano}, \citenamefont {Wang},
  \citenamefont {Vergniory}, \citenamefont {Elcoro}, \citenamefont {Soluyanov},
  \citenamefont {Felser}, \citenamefont {Neupert}, \citenamefont {Regnault},\
  and\ \citenamefont {Bernevig}}]{Wieder2021}%
  \BibitemOpen
  \bibfield  {author} {\bibinfo {author} {\bibfnamefont {B.~J.}\ \bibnamefont
  {Wieder}}, \bibinfo {author} {\bibfnamefont {B.}~\bibnamefont {Bradlyn}},
  \bibinfo {author} {\bibfnamefont {J.}~\bibnamefont {Cano}}, \bibinfo {author}
  {\bibfnamefont {Z.}~\bibnamefont {Wang}}, \bibinfo {author} {\bibfnamefont
  {M.~G.}\ \bibnamefont {Vergniory}}, \bibinfo {author} {\bibfnamefont
  {L.}~\bibnamefont {Elcoro}}, \bibinfo {author} {\bibfnamefont {A.~A.}\
  \bibnamefont {Soluyanov}}, \bibinfo {author} {\bibfnamefont {C.}~\bibnamefont
  {Felser}}, \bibinfo {author} {\bibfnamefont {T.}~\bibnamefont {Neupert}},
  \bibinfo {author} {\bibfnamefont {N.}~\bibnamefont {Regnault}},\ and\
  \bibinfo {author} {\bibfnamefont {B.~A.}\ \bibnamefont {Bernevig}},\
  }\bibfield  {title} {\bibinfo {title} {Topological materials discovery from
  crystal symmetry},\ }\href {https://doi.org/10.1038/s41578-021-00380-2}
  {\bibfield  {journal} {\bibinfo  {journal} {Nat. Rev. Mater.}\ }\textbf
  {\bibinfo {volume} {7}},\ \bibinfo {pages} {196} (\bibinfo {year}
  {2021})}\BibitemShut {NoStop}%
\bibitem [{\citenamefont {Bernevig}\ \emph {et~al.}(2022)\citenamefont
  {Bernevig}, \citenamefont {Felser},\ and\ \citenamefont
  {Beidenkopf}}]{Bernevig2022}%
  \BibitemOpen
  \bibfield  {author} {\bibinfo {author} {\bibfnamefont {B.~A.}\ \bibnamefont
  {Bernevig}}, \bibinfo {author} {\bibfnamefont {C.}~\bibnamefont {Felser}},\
  and\ \bibinfo {author} {\bibfnamefont {H.}~\bibnamefont {Beidenkopf}},\
  }\bibfield  {title} {\bibinfo {title} {Progress and prospects in magnetic
  topological materials},\ }\href {https://doi.org/10.1038/s41586-021-04105-x}
  {\bibfield  {journal} {\bibinfo  {journal} {Nature}\ }\textbf {\bibinfo
  {volume} {603}},\ \bibinfo {pages} {41} (\bibinfo {year} {2022})}\BibitemShut
  {NoStop}%
\bibitem [{\citenamefont {He}\ \emph {et~al.}(2022)\citenamefont {He},
  \citenamefont {Hughes}, \citenamefont {Armitage}, \citenamefont {Tokura},\
  and\ \citenamefont {Wang}}]{He2022}%
  \BibitemOpen
  \bibfield  {author} {\bibinfo {author} {\bibfnamefont {Q.~L.}\ \bibnamefont
  {He}}, \bibinfo {author} {\bibfnamefont {T.~L.}\ \bibnamefont {Hughes}},
  \bibinfo {author} {\bibfnamefont {N.~P.}\ \bibnamefont {Armitage}}, \bibinfo
  {author} {\bibfnamefont {Y.}~\bibnamefont {Tokura}},\ and\ \bibinfo {author}
  {\bibfnamefont {K.~L.}\ \bibnamefont {Wang}},\ }\bibfield  {title} {\bibinfo
  {title} {Topological spintronics and magnetoelectronics},\ }\href
  {https://doi.org/10.1038/s41563-021-01138-5} {\bibfield  {journal} {\bibinfo
  {journal} {Nat. Mater.}\ }\textbf {\bibinfo {volume} {21}},\ \bibinfo {pages}
  {15} (\bibinfo {year} {2022})}\BibitemShut {NoStop}%
\bibitem [{\citenamefont {Yin}\ \emph {et~al.}(2020)\citenamefont {Yin},
  \citenamefont {Ma}, \citenamefont {Cochran}, \citenamefont {Xu},
  \citenamefont {Zhang}, \citenamefont {Tien}, \citenamefont {Shumiya},
  \citenamefont {Cheng}, \citenamefont {Jiang}, \citenamefont {Lian},
  \citenamefont {Song}, \citenamefont {Chang}, \citenamefont {Belopolski},
  \citenamefont {Multer}, \citenamefont {Litskevich}, \citenamefont {Cheng},
  \citenamefont {Yang}, \citenamefont {Swidler}, \citenamefont {Zhou},
  \citenamefont {Lin}, \citenamefont {Neupert}, \citenamefont {Wang},
  \citenamefont {Yao}, \citenamefont {Chang}, \citenamefont {Jia},\ and\
  \citenamefont {Zahid~Hasan}}]{Yin2020}%
  \BibitemOpen
  \bibfield  {author} {\bibinfo {author} {\bibfnamefont {J.-X.}\ \bibnamefont
  {Yin}}, \bibinfo {author} {\bibfnamefont {W.}~\bibnamefont {Ma}}, \bibinfo
  {author} {\bibfnamefont {T.~A.}\ \bibnamefont {Cochran}}, \bibinfo {author}
  {\bibfnamefont {X.}~\bibnamefont {Xu}}, \bibinfo {author} {\bibfnamefont
  {S.~S.}\ \bibnamefont {Zhang}}, \bibinfo {author} {\bibfnamefont {H.-J.}\
  \bibnamefont {Tien}}, \bibinfo {author} {\bibfnamefont {N.}~\bibnamefont
  {Shumiya}}, \bibinfo {author} {\bibfnamefont {G.}~\bibnamefont {Cheng}},
  \bibinfo {author} {\bibfnamefont {K.}~\bibnamefont {Jiang}}, \bibinfo
  {author} {\bibfnamefont {B.}~\bibnamefont {Lian}}, \bibinfo {author}
  {\bibfnamefont {Z.}~\bibnamefont {Song}}, \bibinfo {author} {\bibfnamefont
  {G.}~\bibnamefont {Chang}}, \bibinfo {author} {\bibfnamefont
  {I.}~\bibnamefont {Belopolski}}, \bibinfo {author} {\bibfnamefont
  {D.}~\bibnamefont {Multer}}, \bibinfo {author} {\bibfnamefont
  {M.}~\bibnamefont {Litskevich}}, \bibinfo {author} {\bibfnamefont {Z.-J.}\
  \bibnamefont {Cheng}}, \bibinfo {author} {\bibfnamefont {X.~P.}\ \bibnamefont
  {Yang}}, \bibinfo {author} {\bibfnamefont {B.}~\bibnamefont {Swidler}},
  \bibinfo {author} {\bibfnamefont {H.}~\bibnamefont {Zhou}}, \bibinfo {author}
  {\bibfnamefont {H.}~\bibnamefont {Lin}}, \bibinfo {author} {\bibfnamefont
  {T.}~\bibnamefont {Neupert}}, \bibinfo {author} {\bibfnamefont
  {Z.}~\bibnamefont {Wang}}, \bibinfo {author} {\bibfnamefont {N.}~\bibnamefont
  {Yao}}, \bibinfo {author} {\bibfnamefont {T.-R.}\ \bibnamefont {Chang}},
  \bibinfo {author} {\bibfnamefont {S.}~\bibnamefont {Jia}},\ and\ \bibinfo
  {author} {\bibfnamefont {M.}~\bibnamefont {Zahid~Hasan}},\ }\bibfield
  {title} {\bibinfo {title} {Quantum-limit chern topological magnetism in
  {TbMn$_6$Sn$_6$}},\ }\href {https://doi.org/10.1038/s41586-020-2482-7}
  {\bibfield  {journal} {\bibinfo  {journal} {Nature}\ }\textbf {\bibinfo
  {volume} {583}},\ \bibinfo {pages} {533} (\bibinfo {year}
  {2020})}\BibitemShut {NoStop}%
\bibitem [{\citenamefont {Wei}\ \emph {et~al.}(2025)\citenamefont {Wei},
  \citenamefont {Zhou}, \citenamefont {Tan}, \citenamefont {Gao}, \citenamefont
  {Liu}, \citenamefont {Jing}, \citenamefont {Li}, \citenamefont {Chen},
  \citenamefont {Long}, \citenamefont {Li}, \citenamefont {Qi}, \citenamefont
  {Yan}, \citenamefont {Teng},\ and\ \citenamefont {Chen}}]{Wei2025}%
  \BibitemOpen
  \bibfield  {author} {\bibinfo {author} {\bibfnamefont {Q.}~\bibnamefont
  {Wei}}, \bibinfo {author} {\bibfnamefont {Y.}~\bibnamefont {Zhou}}, \bibinfo
  {author} {\bibfnamefont {H.}~\bibnamefont {Tan}}, \bibinfo {author}
  {\bibfnamefont {L.}~\bibnamefont {Gao}}, \bibinfo {author} {\bibfnamefont
  {R.}~\bibnamefont {Liu}}, \bibinfo {author} {\bibfnamefont {J.}~\bibnamefont
  {Jing}}, \bibinfo {author} {\bibfnamefont {Y.}~\bibnamefont {Li}}, \bibinfo
  {author} {\bibfnamefont {D.}~\bibnamefont {Chen}}, \bibinfo {author}
  {\bibfnamefont {Y.-Z.}\ \bibnamefont {Long}}, \bibinfo {author}
  {\bibfnamefont {Q.}~\bibnamefont {Li}}, \bibinfo {author} {\bibfnamefont
  {Y.}~\bibnamefont {Qi}}, \bibinfo {author} {\bibfnamefont {B.}~\bibnamefont
  {Yan}}, \bibinfo {author} {\bibfnamefont {B.}~\bibnamefont {Teng}},\ and\
  \bibinfo {author} {\bibfnamefont {D.}~\bibnamefont {Chen}},\ }\bibfield
  {title} {\bibinfo {title} {Large anomalous {Hall} effect induced by local
  disorder in the kagome ferrimagnet {TbMn$_6$Sn$_6$}},\ }\href
  {https://doi.org/10.1103/PhysRevB.111.064412} {\bibfield  {journal} {\bibinfo
   {journal} {Phys. Rev. B}\ }\textbf {\bibinfo {volume} {111}},\ \bibinfo
  {pages} {064412} (\bibinfo {year} {2025})}\BibitemShut {NoStop}%
\bibitem [{\citenamefont {Liu}\ \emph {et~al.}(2018{\natexlab{a}})\citenamefont
  {Liu}, \citenamefont {Sun}, \citenamefont {Kumar}, \citenamefont {Muechler},
  \citenamefont {Sun}, \citenamefont {Jiao}, \citenamefont {Yang},
  \citenamefont {Liu}, \citenamefont {Liang}, \citenamefont {Xu}, \citenamefont
  {Kroder}, \citenamefont {S\"{u}\ss{}}, \citenamefont {Borrmann},
  \citenamefont {Shekhar}, \citenamefont {Wang}, \citenamefont {Xi},
  \citenamefont {Wang}, \citenamefont {Schnelle}, \citenamefont {Wirth},
  \citenamefont {Chen}, \citenamefont {Goennenwein},\ and\ \citenamefont
  {Felser}}]{Liu2018a}%
  \BibitemOpen
  \bibfield  {author} {\bibinfo {author} {\bibfnamefont {E.}~\bibnamefont
  {Liu}}, \bibinfo {author} {\bibfnamefont {Y.}~\bibnamefont {Sun}}, \bibinfo
  {author} {\bibfnamefont {N.}~\bibnamefont {Kumar}}, \bibinfo {author}
  {\bibfnamefont {L.}~\bibnamefont {Muechler}}, \bibinfo {author}
  {\bibfnamefont {A.}~\bibnamefont {Sun}}, \bibinfo {author} {\bibfnamefont
  {L.}~\bibnamefont {Jiao}}, \bibinfo {author} {\bibfnamefont {S.-Y.}\
  \bibnamefont {Yang}}, \bibinfo {author} {\bibfnamefont {D.}~\bibnamefont
  {Liu}}, \bibinfo {author} {\bibfnamefont {A.}~\bibnamefont {Liang}}, \bibinfo
  {author} {\bibfnamefont {Q.}~\bibnamefont {Xu}}, \bibinfo {author}
  {\bibfnamefont {J.}~\bibnamefont {Kroder}}, \bibinfo {author} {\bibfnamefont
  {V.}~\bibnamefont {S\"{u}\ss{}}}, \bibinfo {author} {\bibfnamefont
  {H.}~\bibnamefont {Borrmann}}, \bibinfo {author} {\bibfnamefont
  {C.}~\bibnamefont {Shekhar}}, \bibinfo {author} {\bibfnamefont
  {Z.}~\bibnamefont {Wang}}, \bibinfo {author} {\bibfnamefont {C.}~\bibnamefont
  {Xi}}, \bibinfo {author} {\bibfnamefont {W.}~\bibnamefont {Wang}}, \bibinfo
  {author} {\bibfnamefont {W.}~\bibnamefont {Schnelle}}, \bibinfo {author}
  {\bibfnamefont {S.}~\bibnamefont {Wirth}}, \bibinfo {author} {\bibfnamefont
  {Y.}~\bibnamefont {Chen}}, \bibinfo {author} {\bibfnamefont {S.~T.~B.}\
  \bibnamefont {Goennenwein}},\ and\ \bibinfo {author} {\bibfnamefont
  {C.}~\bibnamefont {Felser}},\ }\bibfield  {title} {\bibinfo {title} {Giant
  anomalous {Hall} effect in a ferromagnetic kagome-lattice semimetal},\ }\href
  {https://doi.org/10.1038/s41567-018-0234-5} {\bibfield  {journal} {\bibinfo
  {journal} {Nat. Phys.}\ }\textbf {\bibinfo {volume} {14}},\ \bibinfo {pages}
  {1125} (\bibinfo {year} {2018}{\natexlab{a}})}\BibitemShut {NoStop}%
\bibitem [{\citenamefont {Nayak}\ \emph {et~al.}(2016)\citenamefont {Nayak},
  \citenamefont {Fischer}, \citenamefont {Sun}, \citenamefont {Yan},
  \citenamefont {Karel}, \citenamefont {Komarek}, \citenamefont {Shekhar},
  \citenamefont {Kumar}, \citenamefont {Schnelle}, \citenamefont {Kübler},
  \citenamefont {Felser},\ and\ \citenamefont {Parkin}}]{Nayak2016}%
  \BibitemOpen
  \bibfield  {author} {\bibinfo {author} {\bibfnamefont {A.~K.}\ \bibnamefont
  {Nayak}}, \bibinfo {author} {\bibfnamefont {J.~E.}\ \bibnamefont {Fischer}},
  \bibinfo {author} {\bibfnamefont {Y.}~\bibnamefont {Sun}}, \bibinfo {author}
  {\bibfnamefont {B.}~\bibnamefont {Yan}}, \bibinfo {author} {\bibfnamefont
  {J.}~\bibnamefont {Karel}}, \bibinfo {author} {\bibfnamefont {A.~C.}\
  \bibnamefont {Komarek}}, \bibinfo {author} {\bibfnamefont {C.}~\bibnamefont
  {Shekhar}}, \bibinfo {author} {\bibfnamefont {N.}~\bibnamefont {Kumar}},
  \bibinfo {author} {\bibfnamefont {W.}~\bibnamefont {Schnelle}}, \bibinfo
  {author} {\bibfnamefont {J.}~\bibnamefont {Kübler}}, \bibinfo {author}
  {\bibfnamefont {C.}~\bibnamefont {Felser}},\ and\ \bibinfo {author}
  {\bibfnamefont {S.~S.~P.}\ \bibnamefont {Parkin}},\ }\bibfield  {title}
  {\bibinfo {title} {Large anomalous {Hall} effect driven by a nonvanishing
  {Berry} curvature in the noncolinear antiferromagnet
  {Mn}$_{\textrm{3}}${Ge}},\ }\href {https://doi.org/10.1126/sciadv.1501870}
  {\bibfield  {journal} {\bibinfo  {journal} {Sci. Adv.}\ }\textbf {\bibinfo
  {volume} {2}},\ \bibinfo {pages} {e1501870} (\bibinfo {year}
  {2016})}\BibitemShut {NoStop}%
\bibitem [{\citenamefont {Chen}\ \emph {et~al.}(2014)\citenamefont {Chen},
  \citenamefont {Niu},\ and\ \citenamefont {MacDonald}}]{MacDonald2014}%
  \BibitemOpen
  \bibfield  {author} {\bibinfo {author} {\bibfnamefont {H.}~\bibnamefont
  {Chen}}, \bibinfo {author} {\bibfnamefont {Q.}~\bibnamefont {Niu}},\ and\
  \bibinfo {author} {\bibfnamefont {A.~H.}\ \bibnamefont {MacDonald}},\
  }\bibfield  {title} {\bibinfo {title} {Anomalous {Hall} effect arising from
  noncollinear antiferromagnetism},\ }\href
  {https://doi.org/10.1103/PhysRevLett.112.017205} {\bibfield  {journal}
  {\bibinfo  {journal} {Phys. Rev. Lett.}\ }\textbf {\bibinfo {volume} {112}},\
  \bibinfo {pages} {017205} (\bibinfo {year} {2014})}\BibitemShut {NoStop}%
\bibitem [{\citenamefont {Chen}\ \emph {et~al.}(2021)\citenamefont {Chen},
  \citenamefont {Tomita}, \citenamefont {Minami}, \citenamefont {Fu},
  \citenamefont {Koretsune}, \citenamefont {Kitatani}, \citenamefont
  {Muhammad}, \citenamefont {Nishio-Hamane}, \citenamefont {Ishii},
  \citenamefont {Ishii}, \citenamefont {Arita},\ and\ \citenamefont
  {Nakatsuji}}]{chen2021}%
  \BibitemOpen
  \bibfield  {author} {\bibinfo {author} {\bibfnamefont {T.}~\bibnamefont
  {Chen}}, \bibinfo {author} {\bibfnamefont {T.}~\bibnamefont {Tomita}},
  \bibinfo {author} {\bibfnamefont {S.}~\bibnamefont {Minami}}, \bibinfo
  {author} {\bibfnamefont {M.}~\bibnamefont {Fu}}, \bibinfo {author}
  {\bibfnamefont {T.}~\bibnamefont {Koretsune}}, \bibinfo {author}
  {\bibfnamefont {M.}~\bibnamefont {Kitatani}}, \bibinfo {author}
  {\bibfnamefont {I.}~\bibnamefont {Muhammad}}, \bibinfo {author}
  {\bibfnamefont {D.}~\bibnamefont {Nishio-Hamane}}, \bibinfo {author}
  {\bibfnamefont {R.}~\bibnamefont {Ishii}}, \bibinfo {author} {\bibfnamefont
  {F.}~\bibnamefont {Ishii}}, \bibinfo {author} {\bibfnamefont
  {R.}~\bibnamefont {Arita}},\ and\ \bibinfo {author} {\bibfnamefont
  {S.}~\bibnamefont {Nakatsuji}},\ }\bibfield  {title} {\bibinfo {title}
  {Anomalous transport due to {Weyl} fermions in the chiral antiferromagnets
  {Mn}$_{\textrm{3}}${$\textit{X}$}, {$\textit{X}$} = {Sn}, {Ge}},\ }\href
  {https://doi.org/10.1038/s41467-020-20838-1} {\bibfield  {journal} {\bibinfo
  {journal} {Nat. Commun.}\ }\textbf {\bibinfo {volume} {12}},\ \bibinfo
  {pages} {572} (\bibinfo {year} {2021})}\BibitemShut {NoStop}%
\bibitem [{\citenamefont {Yang}\ \emph {et~al.}(2020)\citenamefont {Yang},
  \citenamefont {Wang}, \citenamefont {Ortiz}, \citenamefont {Liu},
  \citenamefont {Gayles}, \citenamefont {Derunova}, \citenamefont
  {Gonzalez-Hernandez}, \citenamefont {\v{S}mejkal}, \citenamefont {Chen},
  \citenamefont {Parkin}, \citenamefont {Wilson}, \citenamefont {Toberer},
  \citenamefont {McQueen},\ and\ \citenamefont {Ali}}]{yang2020}%
  \BibitemOpen
  \bibfield  {author} {\bibinfo {author} {\bibfnamefont {S.-Y.}\ \bibnamefont
  {Yang}}, \bibinfo {author} {\bibfnamefont {Y.}~\bibnamefont {Wang}}, \bibinfo
  {author} {\bibfnamefont {B.~R.}\ \bibnamefont {Ortiz}}, \bibinfo {author}
  {\bibfnamefont {D.}~\bibnamefont {Liu}}, \bibinfo {author} {\bibfnamefont
  {J.}~\bibnamefont {Gayles}}, \bibinfo {author} {\bibfnamefont
  {E.}~\bibnamefont {Derunova}}, \bibinfo {author} {\bibfnamefont
  {R.}~\bibnamefont {Gonzalez-Hernandez}}, \bibinfo {author} {\bibfnamefont
  {L.}~\bibnamefont {\v{S}mejkal}}, \bibinfo {author} {\bibfnamefont
  {Y.}~\bibnamefont {Chen}}, \bibinfo {author} {\bibfnamefont {S.~S.~P.}\
  \bibnamefont {Parkin}}, \bibinfo {author} {\bibfnamefont {S.~D.}\
  \bibnamefont {Wilson}}, \bibinfo {author} {\bibfnamefont {E.~S.}\
  \bibnamefont {Toberer}}, \bibinfo {author} {\bibfnamefont {T.}~\bibnamefont
  {McQueen}},\ and\ \bibinfo {author} {\bibfnamefont {M.~N.}\ \bibnamefont
  {Ali}},\ }\bibfield  {title} {\bibinfo {title} {Giant, unconventional
  anomalous {Hall} effect in the metallic frustrated magnet candidate,
  {KV}$_{\textrm{3}}${Sb}$_{\textrm{5}}$},\ }\href
  {https://doi.org/10.1126/sciadv.abb6003} {\bibfield  {journal} {\bibinfo
  {journal} {Sci. Adv.}\ }\textbf {\bibinfo {volume} {6}},\ \bibinfo {pages}
  {eabb6003} (\bibinfo {year} {2020})}\BibitemShut {NoStop}%
\bibitem [{\citenamefont {Wang}\ \emph {et~al.}(2025)\citenamefont {Wang},
  \citenamefont {Li}, \citenamefont {Meng}, \citenamefont {Zhang},
  \citenamefont {Lin}, \citenamefont {Li}, \citenamefont {Zheng}, \citenamefont
  {Xu}, \citenamefont {Shang},\ and\ \citenamefont {Zhan}}]{Wang2025}%
  \BibitemOpen
  \bibfield  {author} {\bibinfo {author} {\bibfnamefont {C.}~\bibnamefont
  {Wang}}, \bibinfo {author} {\bibfnamefont {Z.}~\bibnamefont {Li}}, \bibinfo
  {author} {\bibfnamefont {J.}~\bibnamefont {Meng}}, \bibinfo {author}
  {\bibfnamefont {H.}~\bibnamefont {Zhang}}, \bibinfo {author} {\bibfnamefont
  {H.}~\bibnamefont {Lin}}, \bibinfo {author} {\bibfnamefont {J.}~\bibnamefont
  {Li}}, \bibinfo {author} {\bibfnamefont {K.}~\bibnamefont {Zheng}}, \bibinfo
  {author} {\bibfnamefont {Y.}~\bibnamefont {Xu}}, \bibinfo {author}
  {\bibfnamefont {T.}~\bibnamefont {Shang}},\ and\ \bibinfo {author}
  {\bibfnamefont {Q.}~\bibnamefont {Zhan}},\ }\bibfield  {title} {\bibinfo
  {title} {Anomalous {Hall} effect and rich magnetic phase diagram of
  {Mn}$_{100\ensuremath{-}x}${Rh}$_{x}$ epitaxial films},\ }\href
  {https://doi.org/10.1103/cd8d-w92s} {\bibfield  {journal} {\bibinfo
  {journal} {Phys. Rev. B}\ }\textbf {\bibinfo {volume} {112}},\ \bibinfo
  {pages} {224443} (\bibinfo {year} {2025})}\BibitemShut {NoStop}%
\bibitem [{\citenamefont {Chen}\ \emph {et~al.}(2024)\citenamefont {Chen},
  \citenamefont {Lin}, \citenamefont {Lim},\ and\ \citenamefont
  {Ho}}]{Chen2024}%
  \BibitemOpen
  \bibfield  {author} {\bibinfo {author} {\bibfnamefont {S.}~\bibnamefont
  {Chen}}, \bibinfo {author} {\bibfnamefont {D.~J.~X.}\ \bibnamefont {Lin}},
  \bibinfo {author} {\bibfnamefont {B.~C.}\ \bibnamefont {Lim}},\ and\ \bibinfo
  {author} {\bibfnamefont {P.}~\bibnamefont {Ho}},\ }\bibfield  {title}
  {\bibinfo {title} {Mn-based noncollinear antiferromagnets and altermagnets},\
  }\href {https://doi.org/10.1088/1361-6463/ad632b} {\bibfield  {journal}
  {\bibinfo  {journal} {J. Phys. D: Appl. Phys.}\ }\textbf {\bibinfo {volume}
  {57}},\ \bibinfo {pages} {443001} (\bibinfo {year} {2024})}\BibitemShut
  {NoStop}%
\bibitem [{\citenamefont {Kr\'{e}n}\ \emph {et~al.}(1966)\citenamefont
  {Kr\'{e}n}, \citenamefont {K\'{a}d\'{a}r}, \citenamefont {P\'{a}l},
  \citenamefont {S\'{o}lyom},\ and\ \citenamefont {Szab\'{o}}}]{Kren1966}%
  \BibitemOpen
  \bibfield  {author} {\bibinfo {author} {\bibfnamefont {E.}~\bibnamefont
  {Kr\'{e}n}}, \bibinfo {author} {\bibfnamefont {G.}~\bibnamefont
  {K\'{a}d\'{a}r}}, \bibinfo {author} {\bibfnamefont {L.}~\bibnamefont
  {P\'{a}l}}, \bibinfo {author} {\bibfnamefont {J.}~\bibnamefont
  {S\'{o}lyom}},\ and\ \bibinfo {author} {\bibfnamefont {P.}~\bibnamefont
  {Szab\'{o}}},\ }\bibfield  {title} {\bibinfo {title} {Magnetic structures and
  magnetic transformations in ordered {Mn$_3$(Rh,Pt)} alloys},\ }\href
  {https://doi.org/https://doi.org/10.1016/0031-9163(66)90724-4} {\bibfield
  {journal} {\bibinfo  {journal} {Phys. Lett.}\ }\textbf {\bibinfo {volume}
  {20}},\ \bibinfo {pages} {331} (\bibinfo {year} {1966})}\BibitemShut
  {NoStop}%
\bibitem [{\citenamefont {Long}(1991)}]{Long1991}%
  \BibitemOpen
  \bibfield  {author} {\bibinfo {author} {\bibfnamefont {M.~W.}\ \bibnamefont
  {Long}},\ }\bibfield  {title} {\bibinfo {title} {A new magnetic structure for
  {Mn$_3$Pt}},\ }\href {https://doi.org/10.1088/0953-8984/3/36/017} {\bibfield
  {journal} {\bibinfo  {journal} {J. Phys.: Condens. Matter.}\ }\textbf
  {\bibinfo {volume} {3}},\ \bibinfo {pages} {7091} (\bibinfo {year}
  {1991})}\BibitemShut {NoStop}%
\bibitem [{\citenamefont {Zhang}\ \emph {et~al.}(2017)\citenamefont {Zhang},
  \citenamefont {Sun}, \citenamefont {Yang}, \citenamefont {\v{Z}elezn\'{y}},
  \citenamefont {Parkin}, \citenamefont {Felser},\ and\ \citenamefont
  {Yan}}]{Zhang2017}%
  \BibitemOpen
  \bibfield  {author} {\bibinfo {author} {\bibfnamefont {Y.}~\bibnamefont
  {Zhang}}, \bibinfo {author} {\bibfnamefont {Y.}~\bibnamefont {Sun}}, \bibinfo
  {author} {\bibfnamefont {H.}~\bibnamefont {Yang}}, \bibinfo {author}
  {\bibfnamefont {J.}~\bibnamefont {\v{Z}elezn\'{y}}}, \bibinfo {author}
  {\bibfnamefont {S.~P.~P.}\ \bibnamefont {Parkin}}, \bibinfo {author}
  {\bibfnamefont {C.}~\bibnamefont {Felser}},\ and\ \bibinfo {author}
  {\bibfnamefont {B.}~\bibnamefont {Yan}},\ }\bibfield  {title} {\bibinfo
  {title} {Strong anisotropic anomalous {Hall} effect and spin {Hall} effect in
  the chiral antiferromagnetic compounds {Mn}$_{\textrm{3}}${$\textit{X}$}
  ({$\textit{X}$} = {Ge}, {Sn}, {Ga}, {Ir}, {Rh}, and {Pt})},\ }\href
  {https://doi.org/10.1103/PhysRevB.95.075128} {\bibfield  {journal} {\bibinfo
  {journal} {Phys. Rev. B}\ }\textbf {\bibinfo {volume} {95}},\ \bibinfo
  {pages} {075128} (\bibinfo {year} {2017})}\BibitemShut {NoStop}%
\bibitem [{\citenamefont {Liu}\ \emph {et~al.}(2018{\natexlab{b}})\citenamefont
  {Liu}, \citenamefont {Chen}, \citenamefont {Wang}, \citenamefont {Liu},
  \citenamefont {Wang}, \citenamefont {Feng}, \citenamefont {Yan},
  \citenamefont {Wang}, \citenamefont {Jiang}, \citenamefont {Coey},\ and\
  \citenamefont {MacDonald}}]{liu2018}%
  \BibitemOpen
  \bibfield  {author} {\bibinfo {author} {\bibfnamefont {Z.~Q.}\ \bibnamefont
  {Liu}}, \bibinfo {author} {\bibfnamefont {H.}~\bibnamefont {Chen}}, \bibinfo
  {author} {\bibfnamefont {J.~M.}\ \bibnamefont {Wang}}, \bibinfo {author}
  {\bibfnamefont {J.~H.}\ \bibnamefont {Liu}}, \bibinfo {author} {\bibfnamefont
  {K.}~\bibnamefont {Wang}}, \bibinfo {author} {\bibfnamefont {Z.~X.}\
  \bibnamefont {Feng}}, \bibinfo {author} {\bibfnamefont {H.}~\bibnamefont
  {Yan}}, \bibinfo {author} {\bibfnamefont {X.~R.}\ \bibnamefont {Wang}},
  \bibinfo {author} {\bibfnamefont {C.~B.}\ \bibnamefont {Jiang}}, \bibinfo
  {author} {\bibfnamefont {J.~M.~D.}\ \bibnamefont {Coey}},\ and\ \bibinfo
  {author} {\bibfnamefont {A.~H.}\ \bibnamefont {MacDonald}},\ }\bibfield
  {title} {\bibinfo {title} {Electrical switching of the topological anomalous
  {Hall} effect in a non-collinear antiferromagnet above room temperature},\
  }\href {https://doi.org/10.1038/s41928-018-0040-1} {\bibfield  {journal}
  {\bibinfo  {journal} {Nat. Electron.}\ }\textbf {\bibinfo {volume} {1}},\
  \bibinfo {pages} {172} (\bibinfo {year} {2018}{\natexlab{b}})}\BibitemShut
  {NoStop}%
\bibitem [{\citenamefont {Mukherjee}\ \emph {et~al.}(2021)\citenamefont
  {Mukherjee}, \citenamefont {Suraj}, \citenamefont {Basumatary}, \citenamefont
  {Sethupathi},\ and\ \citenamefont {Raman}}]{Mukherjee2021}%
  \BibitemOpen
  \bibfield  {author} {\bibinfo {author} {\bibfnamefont {J.}~\bibnamefont
  {Mukherjee}}, \bibinfo {author} {\bibfnamefont {T.~S.}\ \bibnamefont
  {Suraj}}, \bibinfo {author} {\bibfnamefont {H.}~\bibnamefont {Basumatary}},
  \bibinfo {author} {\bibfnamefont {K.}~\bibnamefont {Sethupathi}},\ and\
  \bibinfo {author} {\bibfnamefont {K.~V.}\ \bibnamefont {Raman}},\ }\bibfield
  {title} {\bibinfo {title} {Sign reversal of anomalous {Hall} conductivity and
  magnetoresistance in cubic noncollinear antiferromagnet {Mn$_3$Pt} thin
  films},\ }\href {https://doi.org/10.1103/PhysRevMaterials.5.014201}
  {\bibfield  {journal} {\bibinfo  {journal} {Phys. Rev. Mater.}\ }\textbf
  {\bibinfo {volume} {5}},\ \bibinfo {pages} {014201} (\bibinfo {year}
  {2021})}\BibitemShut {NoStop}%
\bibitem [{\citenamefont {Zuniga-Cespedes}\ \emph {et~al.}(2023)\citenamefont
  {Zuniga-Cespedes}, \citenamefont {Manna}, \citenamefont {Noad}, \citenamefont
  {Yang}, \citenamefont {Nicklas}, \citenamefont {Felser}, \citenamefont
  {Mackenzie},\ and\ \citenamefont {Hicks}}]{Zuniga2023}%
  \BibitemOpen
  \bibfield  {author} {\bibinfo {author} {\bibfnamefont {B.~E.}\ \bibnamefont
  {Zuniga-Cespedes}}, \bibinfo {author} {\bibfnamefont {K.}~\bibnamefont
  {Manna}}, \bibinfo {author} {\bibfnamefont {H.~M.~L.}\ \bibnamefont {Noad}},
  \bibinfo {author} {\bibfnamefont {P.-Y.}\ \bibnamefont {Yang}}, \bibinfo
  {author} {\bibfnamefont {M.}~\bibnamefont {Nicklas}}, \bibinfo {author}
  {\bibfnamefont {C.}~\bibnamefont {Felser}}, \bibinfo {author} {\bibfnamefont
  {A.~P.}\ \bibnamefont {Mackenzie}},\ and\ \bibinfo {author} {\bibfnamefont
  {C.~W.}\ \bibnamefont {Hicks}},\ }\bibfield  {title} {\bibinfo {title}
  {Observation of an anomalous {Hall} effect in single-crystal {Mn$_3$Pt}},\
  }\href {https://doi.org/10.1088/1367-2630/acbc3f} {\bibfield  {journal}
  {\bibinfo  {journal} {New J. Phys.}\ }\textbf {\bibinfo {volume} {25}},\
  \bibinfo {pages} {023029} (\bibinfo {year} {2023})}\BibitemShut {NoStop}%
\bibitem [{\citenamefont {Chen}\ \emph {et~al.}(2025)\citenamefont {Chen},
  \citenamefont {Lim}, \citenamefont {Lin}, \citenamefont {Soh}, \citenamefont
  {Tan}, \citenamefont {Tan}, \citenamefont {Hnin}, \citenamefont {Wong},
  \citenamefont {Zhang}, \citenamefont {Laskowski}, \citenamefont {Zhao},
  \citenamefont {Chen}, \citenamefont {Khoo},\ and\ \citenamefont
  {Ho}}]{Chen2025}%
  \BibitemOpen
  \bibfield  {author} {\bibinfo {author} {\bibfnamefont {S.}~\bibnamefont
  {Chen}}, \bibinfo {author} {\bibfnamefont {B.~C.}\ \bibnamefont {Lim}},
  \bibinfo {author} {\bibfnamefont {D.~J.~X.}\ \bibnamefont {Lin}}, \bibinfo
  {author} {\bibfnamefont {J.~R.}\ \bibnamefont {Soh}}, \bibinfo {author}
  {\bibfnamefont {H.~R.}\ \bibnamefont {Tan}}, \bibinfo {author} {\bibfnamefont
  {H.~K.}\ \bibnamefont {Tan}}, \bibinfo {author} {\bibfnamefont {Y.~Y.~K.}\
  \bibnamefont {Hnin}}, \bibinfo {author} {\bibfnamefont {S.~K.}\ \bibnamefont
  {Wong}}, \bibinfo {author} {\bibfnamefont {M.}~\bibnamefont {Zhang}},
  \bibinfo {author} {\bibfnamefont {R.}~\bibnamefont {Laskowski}}, \bibinfo
  {author} {\bibfnamefont {T.}~\bibnamefont {Zhao}}, \bibinfo {author}
  {\bibfnamefont {J.}~\bibnamefont {Chen}}, \bibinfo {author} {\bibfnamefont
  {K.~H.}\ \bibnamefont {Khoo}},\ and\ \bibinfo {author} {\bibfnamefont
  {P.}~\bibnamefont {Ho}},\ }\bibfield  {title} {\bibinfo {title} {Tailoring
  antiferromagnetic orders and spin transport in noncollinear
  {Mn}$_{\textrm{3}}${Pt} multilayers},\ }\href
  {https://doi.org/10.1002/adfm.202507406} {\bibfield  {journal} {\bibinfo
  {journal} {Adv. Funct. Mater.}\ }\textbf {\bibinfo {volume} {35}},\ \bibinfo
  {pages} {e07406} (\bibinfo {year} {2025})}\BibitemShut {NoStop}%
\bibitem [{\citenamefont {Xu}\ \emph {et~al.}(2024)\citenamefont {Xu},
  \citenamefont {Dai}, \citenamefont {Jiang}, \citenamefont {Xiong},
  \citenamefont {Cheng}, \citenamefont {Tai}, \citenamefont {Tang},
  \citenamefont {Sun}, \citenamefont {He}, \citenamefont {Yang}, \citenamefont
  {Peng}, \citenamefont {Wang},\ and\ \citenamefont {Zhao}}]{Xu2024}%
  \BibitemOpen
  \bibfield  {author} {\bibinfo {author} {\bibfnamefont {S.}~\bibnamefont
  {Xu}}, \bibinfo {author} {\bibfnamefont {B.}~\bibnamefont {Dai}}, \bibinfo
  {author} {\bibfnamefont {Y.}~\bibnamefont {Jiang}}, \bibinfo {author}
  {\bibfnamefont {D.}~\bibnamefont {Xiong}}, \bibinfo {author} {\bibfnamefont
  {H.}~\bibnamefont {Cheng}}, \bibinfo {author} {\bibfnamefont
  {L.}~\bibnamefont {Tai}}, \bibinfo {author} {\bibfnamefont {M.}~\bibnamefont
  {Tang}}, \bibinfo {author} {\bibfnamefont {Y.}~\bibnamefont {Sun}}, \bibinfo
  {author} {\bibfnamefont {Y.}~\bibnamefont {He}}, \bibinfo {author}
  {\bibfnamefont {B.}~\bibnamefont {Yang}}, \bibinfo {author} {\bibfnamefont
  {Y.}~\bibnamefont {Peng}}, \bibinfo {author} {\bibfnamefont {K.~L.}\
  \bibnamefont {Wang}},\ and\ \bibinfo {author} {\bibfnamefont
  {W.}~\bibnamefont {Zhao}},\ }\bibfield  {title} {\bibinfo {title} {Universal
  scaling law for chiral antiferromagnetism},\ }\href
  {https://doi.org/10.1038/s41467-024-46325-5} {\bibfield  {journal} {\bibinfo
  {journal} {Nat. Commun.}\ }\textbf {\bibinfo {volume} {15}},\ \bibinfo
  {pages} {3717} (\bibinfo {year} {2024})}\BibitemShut {NoStop}%
\bibitem [{\citenamefont {Zhao}\ \emph {et~al.}(2022)\citenamefont {Zhao},
  \citenamefont {Zhang}, \citenamefont {Guo},\ and\ \citenamefont
  {Jiang}}]{Zhao2022}%
  \BibitemOpen
  \bibfield  {author} {\bibinfo {author} {\bibfnamefont {Z.}~\bibnamefont
  {Zhao}}, \bibinfo {author} {\bibfnamefont {K.}~\bibnamefont {Zhang}},
  \bibinfo {author} {\bibfnamefont {Q.}~\bibnamefont {Guo}},\ and\ \bibinfo
  {author} {\bibfnamefont {Y.}~\bibnamefont {Jiang}},\ }\bibfield  {title}
  {\bibinfo {title} {Strain-dependent magnetism and anomalous {Hall} effect in
  noncollinear antiferromagnetic {Mn}$_3${Pt} films},\ }\href
  {https://doi.org/10.1016/j.physe.2022.115141} {\bibfield  {journal} {\bibinfo
   {journal} {Physica E}\ }\textbf {\bibinfo {volume} {138}},\ \bibinfo {pages}
  {115141} (\bibinfo {year} {2022})}\BibitemShut {NoStop}%
\bibitem [{\citenamefont {Novakov}\ \emph {et~al.}(2023)\citenamefont
  {Novakov}, \citenamefont {Meisenheimer}, \citenamefont {Pan}, \citenamefont
  {Kezer}, \citenamefont {Vu}, \citenamefont {Grutter}, \citenamefont {Need},
  \citenamefont {Mundy},\ and\ \citenamefont {Heron}}]{Novakov2023}%
  \BibitemOpen
  \bibfield  {author} {\bibinfo {author} {\bibfnamefont {S.}~\bibnamefont
  {Novakov}}, \bibinfo {author} {\bibfnamefont {P.~B.}\ \bibnamefont
  {Meisenheimer}}, \bibinfo {author} {\bibfnamefont {G.~A.}\ \bibnamefont
  {Pan}}, \bibinfo {author} {\bibfnamefont {P.}~\bibnamefont {Kezer}}, \bibinfo
  {author} {\bibfnamefont {N.~M.}\ \bibnamefont {Vu}}, \bibinfo {author}
  {\bibfnamefont {A.~J.}\ \bibnamefont {Grutter}}, \bibinfo {author}
  {\bibfnamefont {R.~F.}\ \bibnamefont {Need}}, \bibinfo {author}
  {\bibfnamefont {J.~A.}\ \bibnamefont {Mundy}},\ and\ \bibinfo {author}
  {\bibfnamefont {J.~T.}\ \bibnamefont {Heron}},\ }\bibfield  {title} {\bibinfo
  {title} {Composite spin hall conductivity from non-collinear
  antiferromagnetic order},\ }\href {https://doi.org/10.1002/adma.202209866}
  {\bibfield  {journal} {\bibinfo  {journal} {Adv. Mater.}\ }\textbf {\bibinfo
  {volume} {35}},\ \bibinfo {pages} {2209866} (\bibinfo {year}
  {2023})}\BibitemShut {NoStop}%
\bibitem [{\citenamefont {Sinha}\ \emph {et~al.}(2025)\citenamefont {Sinha},
  \citenamefont {Sachin}, \citenamefont {Sinha}, \citenamefont {Roy},
  \citenamefont {Kanungo},\ and\ \citenamefont {Manna}}]{Sinha2025}%
  \BibitemOpen
  \bibfield  {author} {\bibinfo {author} {\bibfnamefont {I.}~\bibnamefont
  {Sinha}}, \bibinfo {author} {\bibfnamefont {S.}~\bibnamefont {Sachin}},
  \bibinfo {author} {\bibfnamefont {S.}~\bibnamefont {Sinha}}, \bibinfo
  {author} {\bibfnamefont {R.}~\bibnamefont {Roy}}, \bibinfo {author}
  {\bibfnamefont {S.}~\bibnamefont {Kanungo}},\ and\ \bibinfo {author}
  {\bibfnamefont {S.}~\bibnamefont {Manna}},\ }\bibfield  {title} {\bibinfo
  {title} {Occurrence of chemically tuned, spin-texture-controlled large
  intrinsic anomalous {Hall} effect in epitaxial {Mn$_{3+x}$Pt$_{1-x}$} thin
  films},\ }\href {https://doi.org/10.1103/hdf1-7vfk} {\bibfield  {journal}
  {\bibinfo  {journal} {Phys. Rev. Mater.}\ }\textbf {\bibinfo {volume} {9}},\
  \bibinfo {pages} {074202} (\bibinfo {year} {2025})}\BibitemShut {NoStop}%
\bibitem [{\citenamefont {An}\ \emph {et~al.}(2020)\citenamefont {An},
  \citenamefont {Tang}, \citenamefont {Hu}, \citenamefont {Yang}, \citenamefont
  {Fan}, \citenamefont {Zhou},\ and\ \citenamefont {Qiu}}]{An2020}%
  \BibitemOpen
  \bibfield  {author} {\bibinfo {author} {\bibfnamefont {N.}~\bibnamefont
  {An}}, \bibinfo {author} {\bibfnamefont {M.}~\bibnamefont {Tang}}, \bibinfo
  {author} {\bibfnamefont {S.}~\bibnamefont {Hu}}, \bibinfo {author}
  {\bibfnamefont {H.}~\bibnamefont {Yang}}, \bibinfo {author} {\bibfnamefont
  {W.}~\bibnamefont {Fan}}, \bibinfo {author} {\bibfnamefont {S.}~\bibnamefont
  {Zhou}},\ and\ \bibinfo {author} {\bibfnamefont {X.}~\bibnamefont {Qiu}},\
  }\bibfield  {title} {\bibinfo {title} {Structure and strain tunings of
  topological anomalous {Hall} effect in cubic noncollinear antiferromagnet
  {Mn$_3$Pt} epitaxial films},\ }\href
  {https://doi.org/10.1007/s11433-019-1525-6} {\bibfield  {journal} {\bibinfo
  {journal} {Sci. China Phys. Mech.}\ }\textbf {\bibinfo {volume} {63}},\
  \bibinfo {pages} {297511} (\bibinfo {year} {2020})}\BibitemShut {NoStop}%
\bibitem [{\citenamefont {Cao}\ \emph {et~al.}(2023)\citenamefont {Cao},
  \citenamefont {Chen}, \citenamefont {Xiao}, \citenamefont {Zhu},
  \citenamefont {Yu}, \citenamefont {Wang}, \citenamefont {Qiu}, \citenamefont
  {Liu}, \citenamefont {Zhao}, \citenamefont {Shao}, \citenamefont {Xu},
  \citenamefont {Chen},\ and\ \citenamefont {Zhan}}]{Cao2023}%
  \BibitemOpen
  \bibfield  {author} {\bibinfo {author} {\bibfnamefont {C.}~\bibnamefont
  {Cao}}, \bibinfo {author} {\bibfnamefont {S.}~\bibnamefont {Chen}}, \bibinfo
  {author} {\bibfnamefont {R.-C.}\ \bibnamefont {Xiao}}, \bibinfo {author}
  {\bibfnamefont {Z.}~\bibnamefont {Zhu}}, \bibinfo {author} {\bibfnamefont
  {G.}~\bibnamefont {Yu}}, \bibinfo {author} {\bibfnamefont {Y.}~\bibnamefont
  {Wang}}, \bibinfo {author} {\bibfnamefont {X.}~\bibnamefont {Qiu}}, \bibinfo
  {author} {\bibfnamefont {L.}~\bibnamefont {Liu}}, \bibinfo {author}
  {\bibfnamefont {T.}~\bibnamefont {Zhao}}, \bibinfo {author} {\bibfnamefont
  {D.-F.}\ \bibnamefont {Shao}}, \bibinfo {author} {\bibfnamefont
  {Y.}~\bibnamefont {Xu}}, \bibinfo {author} {\bibfnamefont {J.}~\bibnamefont
  {Chen}},\ and\ \bibinfo {author} {\bibfnamefont {Q.}~\bibnamefont {Zhan}},\
  }\bibfield  {title} {\bibinfo {title} {Anomalous spin current anisotropy in a
  noncollinear antiferromagnet},\ }\href
  {https://doi.org/10.1038/s41467-023-41568-0} {\bibfield  {journal} {\bibinfo
  {journal} {Nat. Commun.}\ }\textbf {\bibinfo {volume} {14}},\ \bibinfo
  {pages} {5873} (\bibinfo {year} {2023})}\BibitemShut {NoStop}%
\bibitem [{\citenamefont {Qin}\ \emph {et~al.}(2023)\citenamefont {Qin},
  \citenamefont {Yan}, \citenamefont {Wang}, \citenamefont {Chen},
  \citenamefont {Meng}, \citenamefont {Dong}, \citenamefont {Zhu},
  \citenamefont {Cai}, \citenamefont {Feng}, \citenamefont {Zhou},
  \citenamefont {Liu}, \citenamefont {Zhang}, \citenamefont {Zeng},
  \citenamefont {Zhang}, \citenamefont {Jiang},\ and\ \citenamefont
  {Liu}}]{Qin2023}%
  \BibitemOpen
  \bibfield  {author} {\bibinfo {author} {\bibfnamefont {P.}~\bibnamefont
  {Qin}}, \bibinfo {author} {\bibfnamefont {H.}~\bibnamefont {Yan}}, \bibinfo
  {author} {\bibfnamefont {X.}~\bibnamefont {Wang}}, \bibinfo {author}
  {\bibfnamefont {H.}~\bibnamefont {Chen}}, \bibinfo {author} {\bibfnamefont
  {Z.}~\bibnamefont {Meng}}, \bibinfo {author} {\bibfnamefont {J.}~\bibnamefont
  {Dong}}, \bibinfo {author} {\bibfnamefont {M.}~\bibnamefont {Zhu}}, \bibinfo
  {author} {\bibfnamefont {J.}~\bibnamefont {Cai}}, \bibinfo {author}
  {\bibfnamefont {Z.}~\bibnamefont {Feng}}, \bibinfo {author} {\bibfnamefont
  {X.}~\bibnamefont {Zhou}}, \bibinfo {author} {\bibfnamefont {L.}~\bibnamefont
  {Liu}}, \bibinfo {author} {\bibfnamefont {T.}~\bibnamefont {Zhang}}, \bibinfo
  {author} {\bibfnamefont {Z.}~\bibnamefont {Zeng}}, \bibinfo {author}
  {\bibfnamefont {J.}~\bibnamefont {Zhang}}, \bibinfo {author} {\bibfnamefont
  {C.}~\bibnamefont {Jiang}},\ and\ \bibinfo {author} {\bibfnamefont
  {Z.}~\bibnamefont {Liu}},\ }\bibfield  {title} {\bibinfo {title}
  {Room-temperature magnetoresistance in an all-antiferromagnetic tunnel
  junction},\ }\href {https://doi.org/10.1038/s41586-022-05461-y} {\bibfield
  {journal} {\bibinfo  {journal} {Nature}\ }\textbf {\bibinfo {volume} {613}},\
  \bibinfo {pages} {485} (\bibinfo {year} {2023})}\BibitemShut {NoStop}%
\bibitem [{\citenamefont {Kr\'{e}n}\ \emph {et~al.}(1968)\citenamefont
  {Kr\'{e}n}, \citenamefont {K\'{a}d\'{a}r}, \citenamefont {P\'{a}l},
  \citenamefont {S\'{o}lyom}, \citenamefont {Szab\'{o}},\ and\ \citenamefont
  {Tarn\'{o}czi}}]{Kren1968}%
  \BibitemOpen
  \bibfield  {author} {\bibinfo {author} {\bibfnamefont {E.}~\bibnamefont
  {Kr\'{e}n}}, \bibinfo {author} {\bibfnamefont {G.}~\bibnamefont
  {K\'{a}d\'{a}r}}, \bibinfo {author} {\bibfnamefont {L.}~\bibnamefont
  {P\'{a}l}}, \bibinfo {author} {\bibfnamefont {J.}~\bibnamefont {S\'{o}lyom}},
  \bibinfo {author} {\bibfnamefont {P.}~\bibnamefont {Szab\'{o}}},\ and\
  \bibinfo {author} {\bibfnamefont {T.}~\bibnamefont {Tarn\'{o}czi}},\
  }\bibfield  {title} {\bibinfo {title} {Magnetic structures and exchange
  interactions in the {Mn-Pt} system},\ }\href
  {https://doi.org/10.1103/PhysRev.171.574} {\bibfield  {journal} {\bibinfo
  {journal} {Phys. Rev.}\ }\textbf {\bibinfo {volume} {171}},\ \bibinfo {pages}
  {574} (\bibinfo {year} {1968})}\BibitemShut {NoStop}%
\bibitem [{\citenamefont {Parkin}\ \emph {et~al.}(2003)\citenamefont {Parkin},
  \citenamefont {{Xin Jiang}}, \citenamefont {Kaiser}, \citenamefont
  {Panchula}, \citenamefont {Roche},\ and\ \citenamefont
  {Samant}}]{Parkin2003}%
  \BibitemOpen
  \bibfield  {author} {\bibinfo {author} {\bibfnamefont {S.}~\bibnamefont
  {Parkin}}, \bibinfo {author} {\bibnamefont {{Xin Jiang}}}, \bibinfo {author}
  {\bibfnamefont {C.}~\bibnamefont {Kaiser}}, \bibinfo {author} {\bibfnamefont
  {A.}~\bibnamefont {Panchula}}, \bibinfo {author} {\bibfnamefont
  {K.}~\bibnamefont {Roche}},\ and\ \bibinfo {author} {\bibfnamefont
  {M.}~\bibnamefont {Samant}},\ }\bibfield  {title} {\bibinfo {title}
  {Magnetically engineered spintronic sensors and memory},\ }\href
  {https://doi.org/10.1109/JPROC.2003.811807} {\bibfield  {journal} {\bibinfo
  {journal} {Proc. IEEE}\ }\textbf {\bibinfo {volume} {91}},\ \bibinfo {pages}
  {661} (\bibinfo {year} {2003})}\BibitemShut {NoStop}%
\bibitem [{\citenamefont {Iusipova}\ and\ \citenamefont
  {Popov}(2021)}]{Iusipova2021}%
  \BibitemOpen
  \bibfield  {author} {\bibinfo {author} {\bibfnamefont {I.~A.}\ \bibnamefont
  {Iusipova}}\ and\ \bibinfo {author} {\bibfnamefont {A.~I.}\ \bibnamefont
  {Popov}},\ }\bibfield  {title} {\bibinfo {title} {Spin valves in
  microelectronics ({A} review)},\ }\href
  {https://doi.org/10.1134/S1063782621130108} {\bibfield  {journal} {\bibinfo
  {journal} {Semiconductors}\ }\textbf {\bibinfo {volume} {55}},\ \bibinfo
  {pages} {1008} (\bibinfo {year} {2021})}\BibitemShut {NoStop}%
\bibitem [{\citenamefont {Antonini}\ \emph {et~al.}(1969)\citenamefont
  {Antonini}, \citenamefont {Lucari}, \citenamefont {Menzinger},\ and\
  \citenamefont {Paoletti}}]{Antonini1969}%
  \BibitemOpen
  \bibfield  {author} {\bibinfo {author} {\bibfnamefont {B.}~\bibnamefont
  {Antonini}}, \bibinfo {author} {\bibfnamefont {F.}~\bibnamefont {Lucari}},
  \bibinfo {author} {\bibfnamefont {F.}~\bibnamefont {Menzinger}},\ and\
  \bibinfo {author} {\bibfnamefont {A.}~\bibnamefont {Paoletti}},\ }\bibfield
  {title} {\bibinfo {title} {Magnetization distribution in ferromagnetic
  {Mn}{Pt}$_3$ by a polarized-neutron investigation},\ }\href
  {https://doi.org/10.1103/PhysRev.187.611} {\bibfield  {journal} {\bibinfo
  {journal} {Phys. Rev.}\ }\textbf {\bibinfo {volume} {187}},\ \bibinfo {pages}
  {611} (\bibinfo {year} {1969})}\BibitemShut {NoStop}%
\bibitem [{\citenamefont {Wierman}\ \emph {et~al.}(1997)\citenamefont
  {Wierman}, \citenamefont {Hilfiker}, \citenamefont {Sabiryanov},
  \citenamefont {Jaswal}, \citenamefont {Kirby},\ and\ \citenamefont
  {Woollam}}]{Wierman1997}%
  \BibitemOpen
  \bibfield  {author} {\bibinfo {author} {\bibfnamefont {K.~W.}\ \bibnamefont
  {Wierman}}, \bibinfo {author} {\bibfnamefont {J.~N.}\ \bibnamefont
  {Hilfiker}}, \bibinfo {author} {\bibfnamefont {R.~F.}\ \bibnamefont
  {Sabiryanov}}, \bibinfo {author} {\bibfnamefont {S.~S.}\ \bibnamefont
  {Jaswal}}, \bibinfo {author} {\bibfnamefont {R.~D.}\ \bibnamefont {Kirby}},\
  and\ \bibinfo {author} {\bibfnamefont {J.~A.}\ \bibnamefont {Woollam}},\
  }\bibfield  {title} {\bibinfo {title} {Optical and magneto-optical constants
  of {Mn}{Pt}$_3$},\ }\href {https://doi.org/10.1103/PhysRevB.55.3093}
  {\bibfield  {journal} {\bibinfo  {journal} {Phys. Rev. B}\ }\textbf {\bibinfo
  {volume} {55}},\ \bibinfo {pages} {3093} (\bibinfo {year}
  {1997})}\BibitemShut {NoStop}%
\bibitem [{\citenamefont {Kato}\ \emph {et~al.}(1995)\citenamefont {Kato},
  \citenamefont {Kikuzawa}, \citenamefont {Iwata}, \citenamefont {Tsunashima},\
  and\ \citenamefont {Uchiyama}}]{Kato1995}%
  \BibitemOpen
  \bibfield  {author} {\bibinfo {author} {\bibfnamefont {T.}~\bibnamefont
  {Kato}}, \bibinfo {author} {\bibfnamefont {H.}~\bibnamefont {Kikuzawa}},
  \bibinfo {author} {\bibfnamefont {S.}~\bibnamefont {Iwata}}, \bibinfo
  {author} {\bibfnamefont {S.}~\bibnamefont {Tsunashima}},\ and\ \bibinfo
  {author} {\bibfnamefont {S.}~\bibnamefont {Uchiyama}},\ }\bibfield  {title}
  {\bibinfo {title} {Magneto-optical effect in {MnP}t$_3$ alloy films},\ }\href
  {https://doi.org/10.1016/0304-8853(94)01572-4} {\bibfield  {journal}
  {\bibinfo  {journal} {J. Magn. Magn. Mater.}\ }\textbf {\bibinfo {volume}
  {140-144}},\ \bibinfo {pages} {713} (\bibinfo {year} {1995})}\BibitemShut
  {NoStop}%
\bibitem [{\citenamefont {Oppeneer}\ \emph {et~al.}(1996)\citenamefont
  {Oppeneer}, \citenamefont {Antonov}, \citenamefont {Kraft}, \citenamefont
  {Eschrig}, \citenamefont {Yaresko},\ and\ \citenamefont
  {Perlov}}]{Oppeneer1996}%
  \BibitemOpen
  \bibfield  {author} {\bibinfo {author} {\bibfnamefont {P.~M.}\ \bibnamefont
  {Oppeneer}}, \bibinfo {author} {\bibfnamefont {V.~N.}\ \bibnamefont
  {Antonov}}, \bibinfo {author} {\bibfnamefont {T.}~\bibnamefont {Kraft}},
  \bibinfo {author} {\bibfnamefont {H.}~\bibnamefont {Eschrig}}, \bibinfo
  {author} {\bibfnamefont {A.~N.}\ \bibnamefont {Yaresko}},\ and\ \bibinfo
  {author} {\bibfnamefont {A.~Y.}\ \bibnamefont {Perlov}},\ }\bibfield  {title}
  {\bibinfo {title} {Calculated magneto-optical {Kerr} spectra of
  {$\textit{X}$Pt$_3$} compounds ({$\textit{X}$} = {V}, {Cr}, {Mn}, {Fe} and
  {Co})},\ }\href {https://doi.org/10.1088/0953-8984/8/31/010} {\bibfield
  {journal} {\bibinfo  {journal} {J. Phys.: Condens. Matter}\ }\textbf
  {\bibinfo {volume} {8}},\ \bibinfo {pages} {5769} (\bibinfo {year}
  {1996})}\BibitemShut {NoStop}%
\bibitem [{\citenamefont {Markou}\ \emph {et~al.}(2021)\citenamefont {Markou},
  \citenamefont {Gayles}, \citenamefont {Derunova}, \citenamefont {Swekis},
  \citenamefont {Noky}, \citenamefont {Zhang}, \citenamefont {Ali},
  \citenamefont {Sun},\ and\ \citenamefont {Felser}}]{Markou2021}%
  \BibitemOpen
  \bibfield  {author} {\bibinfo {author} {\bibfnamefont {A.}~\bibnamefont
  {Markou}}, \bibinfo {author} {\bibfnamefont {J.}~\bibnamefont {Gayles}},
  \bibinfo {author} {\bibfnamefont {E.}~\bibnamefont {Derunova}}, \bibinfo
  {author} {\bibfnamefont {P.}~\bibnamefont {Swekis}}, \bibinfo {author}
  {\bibfnamefont {J.}~\bibnamefont {Noky}}, \bibinfo {author} {\bibfnamefont
  {L.}~\bibnamefont {Zhang}}, \bibinfo {author} {\bibfnamefont {M.~N.}\
  \bibnamefont {Ali}}, \bibinfo {author} {\bibfnamefont {Y.}~\bibnamefont
  {Sun}},\ and\ \bibinfo {author} {\bibfnamefont {C.}~\bibnamefont {Felser}},\
  }\bibfield  {title} {\bibinfo {title} {Hard magnet topological semimetals in
  {$X$Pt$_3$} compounds with the harmony of berry curvature},\ }\href
  {https://doi.org/10.1038/s42005-021-00608-1} {\bibfield  {journal} {\bibinfo
  {journal} {Commun. Phys.}\ }\textbf {\bibinfo {volume} {4}},\ \bibinfo
  {pages} {104} (\bibinfo {year} {2021})}\BibitemShut {NoStop}%
\bibitem [{Sup()}]{Supple}%
  \BibitemOpen
  \href {https://doi.org/xxx} {}\bibinfo {note} {For details on the
  measurements of crystal structure, electrical resistivity, and magnetization
  of {MnPt$_3$} films with different thicknesses, as well as for the data
  analysis, see the Supplementary Material at
  http://link.aps.org/supplemental/XXX/PhysRevB.XXX}\BibitemShut {NoStop}%
\bibitem [{\citenamefont {Fujishiro}\ \emph {et~al.}(2021)\citenamefont
  {Fujishiro}, \citenamefont {Kanazawa}, \citenamefont {Kurihara},
  \citenamefont {Ishizuka}, \citenamefont {Hori}, \citenamefont {Yasin},
  \citenamefont {Yu}, \citenamefont {Tsukazaki}, \citenamefont {Ichikawa},
  \citenamefont {Kawasaki}, \citenamefont {Nagaosa}, \citenamefont {Tokunaga},\
  and\ \citenamefont {Tokura}}]{Fujishiro2021}%
  \BibitemOpen
  \bibfield  {author} {\bibinfo {author} {\bibfnamefont {Y.}~\bibnamefont
  {Fujishiro}}, \bibinfo {author} {\bibfnamefont {N.}~\bibnamefont {Kanazawa}},
  \bibinfo {author} {\bibfnamefont {R.}~\bibnamefont {Kurihara}}, \bibinfo
  {author} {\bibfnamefont {H.}~\bibnamefont {Ishizuka}}, \bibinfo {author}
  {\bibfnamefont {T.}~\bibnamefont {Hori}}, \bibinfo {author} {\bibfnamefont
  {F.~S.}\ \bibnamefont {Yasin}}, \bibinfo {author} {\bibfnamefont
  {X.}~\bibnamefont {Yu}}, \bibinfo {author} {\bibfnamefont {A.}~\bibnamefont
  {Tsukazaki}}, \bibinfo {author} {\bibfnamefont {M.}~\bibnamefont {Ichikawa}},
  \bibinfo {author} {\bibfnamefont {M.}~\bibnamefont {Kawasaki}}, \bibinfo
  {author} {\bibfnamefont {N.}~\bibnamefont {Nagaosa}}, \bibinfo {author}
  {\bibfnamefont {M.}~\bibnamefont {Tokunaga}},\ and\ \bibinfo {author}
  {\bibfnamefont {Y.}~\bibnamefont {Tokura}},\ }\bibfield  {title} {\bibinfo
  {title} {Giant anomalous {Hall} effect from spin-chirality scattering in a
  chiral magnet},\ }\href {https://doi.org/10.1038/s41467-020-20384-w}
  {\bibfield  {journal} {\bibinfo  {journal} {Nat. Commun.}\ }\textbf {\bibinfo
  {volume} {12}},\ \bibinfo {pages} {317} (\bibinfo {year} {2021})}\BibitemShut
  {NoStop}%
\bibitem [{\citenamefont {Park}\ \emph {et~al.}(2020)\citenamefont {Park},
  \citenamefont {Reichlova}, \citenamefont {Schlitz}, \citenamefont {Lammel},
  \citenamefont {Markou}, \citenamefont {Swekis}, \citenamefont {Ritzinger},
  \citenamefont {Kriegner}, \citenamefont {Noky}, \citenamefont {Gayles},
  \citenamefont {Sun}, \citenamefont {Felser}, \citenamefont {Nielsch},
  \citenamefont {Goennenwein},\ and\ \citenamefont {Thomas}}]{Park2020}%
  \BibitemOpen
  \bibfield  {author} {\bibinfo {author} {\bibfnamefont {G.-H.}\ \bibnamefont
  {Park}}, \bibinfo {author} {\bibfnamefont {H.}~\bibnamefont {Reichlova}},
  \bibinfo {author} {\bibfnamefont {R.}~\bibnamefont {Schlitz}}, \bibinfo
  {author} {\bibfnamefont {M.}~\bibnamefont {Lammel}}, \bibinfo {author}
  {\bibfnamefont {A.}~\bibnamefont {Markou}}, \bibinfo {author} {\bibfnamefont
  {P.}~\bibnamefont {Swekis}}, \bibinfo {author} {\bibfnamefont
  {P.}~\bibnamefont {Ritzinger}}, \bibinfo {author} {\bibfnamefont
  {D.}~\bibnamefont {Kriegner}}, \bibinfo {author} {\bibfnamefont
  {J.}~\bibnamefont {Noky}}, \bibinfo {author} {\bibfnamefont {J.}~\bibnamefont
  {Gayles}}, \bibinfo {author} {\bibfnamefont {Y.}~\bibnamefont {Sun}},
  \bibinfo {author} {\bibfnamefont {C.}~\bibnamefont {Felser}}, \bibinfo
  {author} {\bibfnamefont {K.}~\bibnamefont {Nielsch}}, \bibinfo {author}
  {\bibfnamefont {S.~T.~B.}\ \bibnamefont {Goennenwein}},\ and\ \bibinfo
  {author} {\bibfnamefont {A.}~\bibnamefont {Thomas}},\ }\bibfield  {title}
  {\bibinfo {title} {Thickness dependence of the anomalous nernst effect and
  the mott relation of weyl semimetal ${\mathrm{co}}_{2}\mathrm{MnGa}$ thin
  films},\ }\href {https://doi.org/10.1103/PhysRevB.101.060406} {\bibfield
  {journal} {\bibinfo  {journal} {Phys. Rev. B}\ }\textbf {\bibinfo {volume}
  {101}},\ \bibinfo {pages} {060406} (\bibinfo {year} {2020})}\BibitemShut
  {NoStop}%
\bibitem [{\citenamefont {Zhu}\ \emph {et~al.}(2023)\citenamefont {Zhu},
  \citenamefont {Li}, \citenamefont {Meng}, \citenamefont {Feng}, \citenamefont
  {Zhen}, \citenamefont {Lin}, \citenamefont {Yu}, \citenamefont {Cheng},
  \citenamefont {Jiang}, \citenamefont {Xu}, \citenamefont {Shang},\ and\
  \citenamefont {Zhan}}]{Zhu2023}%
  \BibitemOpen
  \bibfield  {author} {\bibinfo {author} {\bibfnamefont {X.}~\bibnamefont
  {Zhu}}, \bibinfo {author} {\bibfnamefont {H.}~\bibnamefont {Li}}, \bibinfo
  {author} {\bibfnamefont {J.}~\bibnamefont {Meng}}, \bibinfo {author}
  {\bibfnamefont {X.}~\bibnamefont {Feng}}, \bibinfo {author} {\bibfnamefont
  {Z.}~\bibnamefont {Zhen}}, \bibinfo {author} {\bibfnamefont {H.}~\bibnamefont
  {Lin}}, \bibinfo {author} {\bibfnamefont {B.}~\bibnamefont {Yu}}, \bibinfo
  {author} {\bibfnamefont {W.}~\bibnamefont {Cheng}}, \bibinfo {author}
  {\bibfnamefont {D.}~\bibnamefont {Jiang}}, \bibinfo {author} {\bibfnamefont
  {Y.}~\bibnamefont {Xu}}, \bibinfo {author} {\bibfnamefont {T.}~\bibnamefont
  {Shang}},\ and\ \bibinfo {author} {\bibfnamefont {Q.}~\bibnamefont {Zhan}},\
  }\bibfield  {title} {\bibinfo {title} {Absence of topological {Hall} effect
  in {Fe$_x$Rh$_{100-x}$} epitaxial films: {Revisiting} their phase diagram},\
  }\href {https://doi.org/10.1103/PhysRevB.108.144437} {\bibfield  {journal}
  {\bibinfo  {journal} {Phys. Rev. B}\ }\textbf {\bibinfo {volume} {108}},\
  \bibinfo {pages} {144437} (\bibinfo {year} {2023})}\BibitemShut {NoStop}%
\bibitem [{\citenamefont {Wierman}\ and\ \citenamefont
  {Kirby}(1996)}]{Kurt1996}%
  \BibitemOpen
  \bibfield  {author} {\bibinfo {author} {\bibfnamefont {K.~W.}\ \bibnamefont
  {Wierman}}\ and\ \bibinfo {author} {\bibfnamefont {R.~D.}\ \bibnamefont
  {Kirby}},\ }\bibfield  {title} {\bibinfo {title} {Long-range order and
  magnetic properties of {Mn$_x$Pt$_{1-x}$} thin films},\ }\href
  {https://doi.org/Doi 10.1016/0304-8853(95)00581-1} {\bibfield  {journal}
  {\bibinfo  {journal} {J. Magn. Magn. Mater.}\ }\textbf {\bibinfo {volume}
  {154}},\ \bibinfo {pages} {12} (\bibinfo {year} {1996})}\BibitemShut
  {NoStop}%
\bibitem [{\citenamefont {Ishino}\ \emph {et~al.}(2018)\citenamefont {Ishino},
  \citenamefont {So}, \citenamefont {Goto}, \citenamefont {Hajiri},\ and\
  \citenamefont {Asano}}]{Ishino2018}%
  \BibitemOpen
  \bibfield  {author} {\bibinfo {author} {\bibfnamefont {S.}~\bibnamefont
  {Ishino}}, \bibinfo {author} {\bibfnamefont {J.}~\bibnamefont {So}}, \bibinfo
  {author} {\bibfnamefont {H.}~\bibnamefont {Goto}}, \bibinfo {author}
  {\bibfnamefont {T.}~\bibnamefont {Hajiri}},\ and\ \bibinfo {author}
  {\bibfnamefont {H.}~\bibnamefont {Asano}},\ }\bibfield  {title} {\bibinfo
  {title} {Preparation and evaluation of {Mn}$_3${GaN}$_{1-x}$ thin films with
  controlled {N} compositions},\ }\href {https://doi.org/10.1063/1.5007333}
  {\bibfield  {journal} {\bibinfo  {journal} {AIP Advances}\ }\textbf {\bibinfo
  {volume} {8}},\ \bibinfo {pages} {056312} (\bibinfo {year}
  {2018})}\BibitemShut {NoStop}%
\bibitem [{\citenamefont {Li}\ \emph {et~al.}(2025)\citenamefont {Li},
  \citenamefont {Zhou}, \citenamefont {Xu}, \citenamefont {Li}, \citenamefont
  {Yuan}, \citenamefont {Lai}, \citenamefont {Song}, \citenamefont {Liu},
  \citenamefont {Liu}, \citenamefont {Zhang}, \citenamefont {Lu},\ and\
  \citenamefont {Xiong}}]{Li2025}%
  \BibitemOpen
  \bibfield  {author} {\bibinfo {author} {\bibfnamefont {Q.}~\bibnamefont
  {Li}}, \bibinfo {author} {\bibfnamefont {C.}~\bibnamefont {Zhou}}, \bibinfo
  {author} {\bibfnamefont {Y.}~\bibnamefont {Xu}}, \bibinfo {author}
  {\bibfnamefont {R.}~\bibnamefont {Li}}, \bibinfo {author} {\bibfnamefont
  {X.}~\bibnamefont {Yuan}}, \bibinfo {author} {\bibfnamefont {H.}~\bibnamefont
  {Lai}}, \bibinfo {author} {\bibfnamefont {Y.}~\bibnamefont {Song}}, \bibinfo
  {author} {\bibfnamefont {F.}~\bibnamefont {Liu}}, \bibinfo {author}
  {\bibfnamefont {Y.}~\bibnamefont {Liu}}, \bibinfo {author} {\bibfnamefont
  {Z.}~\bibnamefont {Zhang}}, \bibinfo {author} {\bibfnamefont
  {Z.}~\bibnamefont {Lu}},\ and\ \bibinfo {author} {\bibfnamefont
  {R.}~\bibnamefont {Xiong}},\ }\bibfield  {title} {\bibinfo {title} {Enhancing
  large effective spin-torque efficiency in {MnRh} by magnetic phase
  transition},\ }\href {https://doi.org/10.1063/5.0254025} {\bibfield
  {journal} {\bibinfo  {journal} {Appl. Phys. Lett.}\ }\textbf {\bibinfo
  {volume} {126}},\ \bibinfo {pages} {152407} (\bibinfo {year}
  {2025})}\BibitemShut {NoStop}%
\bibitem [{\citenamefont {Kobayashi}\ \emph {et~al.}(2022)\citenamefont
  {Kobayashi}, \citenamefont {Kimata}, \citenamefont {Kan}, \citenamefont
  {Ikebuchi}, \citenamefont {Shiota}, \citenamefont {Kohno}, \citenamefont
  {Shimakawa}, \citenamefont {Ono},\ and\ \citenamefont
  {Moriyama}}]{Kobayashi2022}%
  \BibitemOpen
  \bibfield  {author} {\bibinfo {author} {\bibfnamefont {Y.}~\bibnamefont
  {Kobayashi}}, \bibinfo {author} {\bibfnamefont {M.}~\bibnamefont {Kimata}},
  \bibinfo {author} {\bibfnamefont {D.}~\bibnamefont {Kan}}, \bibinfo {author}
  {\bibfnamefont {T.}~\bibnamefont {Ikebuchi}}, \bibinfo {author}
  {\bibfnamefont {Y.}~\bibnamefont {Shiota}}, \bibinfo {author} {\bibfnamefont
  {H.}~\bibnamefont {Kohno}}, \bibinfo {author} {\bibfnamefont
  {Y.}~\bibnamefont {Shimakawa}}, \bibinfo {author} {\bibfnamefont
  {T.}~\bibnamefont {Ono}},\ and\ \bibinfo {author} {\bibfnamefont
  {T.}~\bibnamefont {Moriyama}},\ }\bibfield  {title} {\bibinfo {title}
  {Extrinsic contribution to anomalous {Hall} effect in chiral
  antiferromagnetic (111)-oriented {$L1_2$}-{Mn}$_{\textrm{3}}${Ir} films},\
  }\href {https://doi.org/10.35848/1347-4065/ac7625} {\bibfield  {journal}
  {\bibinfo  {journal} {Jpn. J. Appl. Phys.}\ }\textbf {\bibinfo {volume}
  {61}},\ \bibinfo {pages} {070912} (\bibinfo {year} {2022})}\BibitemShut
  {NoStop}%
\bibitem [{\citenamefont {Nozi\`{e}res}\ and\ \citenamefont
  {Lewiner}(1973)}]{Lewiner1973}%
  \BibitemOpen
  \bibfield  {author} {\bibinfo {author} {\bibfnamefont {P.}~\bibnamefont
  {Nozi\`{e}res}}\ and\ \bibinfo {author} {\bibfnamefont {C.}~\bibnamefont
  {Lewiner}},\ }\bibfield  {title} {\bibinfo {title} {A simple theory of the
  anomalous {Hall} effect in semiconductors},\ }\href
  {https://doi.org/10.1051/jphys:019730034010090100} {\bibfield  {journal}
  {\bibinfo  {journal} {Journal de Physique}\ }\textbf {\bibinfo {volume}
  {34}},\ \bibinfo {pages} {901} (\bibinfo {year} {1973})}\BibitemShut
  {NoStop}%
\bibitem [{\citenamefont {Onoda}\ \emph {et~al.}(2006)\citenamefont {Onoda},
  \citenamefont {Sugimoto},\ and\ \citenamefont {Nagaosa}}]{Onoda2006}%
  \BibitemOpen
  \bibfield  {author} {\bibinfo {author} {\bibfnamefont {S.}~\bibnamefont
  {Onoda}}, \bibinfo {author} {\bibfnamefont {N.}~\bibnamefont {Sugimoto}},\
  and\ \bibinfo {author} {\bibfnamefont {N.}~\bibnamefont {Nagaosa}},\
  }\bibfield  {title} {\bibinfo {title} {Intrinsic versus extrinsic anomalous
  {Hall} effect in ferromagnets},\ }\href
  {https://doi.org/10.1103/PhysRevLett.97.126602} {\bibfield  {journal}
  {\bibinfo  {journal} {Phys. Rev. Lett.}\ }\textbf {\bibinfo {volume} {97}},\
  \bibinfo {pages} {126602} (\bibinfo {year} {2006})}\BibitemShut {NoStop}%
\bibitem [{\citenamefont {Wang}\ \emph {et~al.}(2016)\citenamefont {Wang},
  \citenamefont {Sun}, \citenamefont {Zhang}, \citenamefont {Pang},\ and\
  \citenamefont {Lei}}]{Wang2016}%
  \BibitemOpen
  \bibfield  {author} {\bibinfo {author} {\bibfnamefont {Q.}~\bibnamefont
  {Wang}}, \bibinfo {author} {\bibfnamefont {S.}~\bibnamefont {Sun}}, \bibinfo
  {author} {\bibfnamefont {X.}~\bibnamefont {Zhang}}, \bibinfo {author}
  {\bibfnamefont {F.}~\bibnamefont {Pang}},\ and\ \bibinfo {author}
  {\bibfnamefont {H.}~\bibnamefont {Lei}},\ }\bibfield  {title} {\bibinfo
  {title} {Anomalous {Hall} effect in a ferromagnetic {Fe$_3$Sn$_2$} single
  crystal with a geometrically frustrated {Fe} bilayer kagome lattice},\ }\href
  {https://doi.org/10.1103/PhysRevB.94.075135} {\bibfield  {journal} {\bibinfo
  {journal} {Phys. Rev. B}\ }\textbf {\bibinfo {volume} {94}},\ \bibinfo
  {pages} {075135} (\bibinfo {year} {2016})}\BibitemShut {NoStop}%
\bibitem [{\citenamefont {Kim}\ \emph {et~al.}(2018)\citenamefont {Kim},
  \citenamefont {Seo}, \citenamefont {Lee}, \citenamefont {Ko}, \citenamefont
  {Kim}, \citenamefont {Jang}, \citenamefont {Ok}, \citenamefont {Lee},
  \citenamefont {Jo}, \citenamefont {Kang}, \citenamefont {Shim}, \citenamefont
  {Kim}, \citenamefont {Yeom}, \citenamefont {Il~Min}, \citenamefont {Yang},\
  and\ \citenamefont {Kim}}]{Kim2018}%
  \BibitemOpen
  \bibfield  {author} {\bibinfo {author} {\bibfnamefont {K.}~\bibnamefont
  {Kim}}, \bibinfo {author} {\bibfnamefont {J.}~\bibnamefont {Seo}}, \bibinfo
  {author} {\bibfnamefont {E.}~\bibnamefont {Lee}}, \bibinfo {author}
  {\bibfnamefont {K.-T.}\ \bibnamefont {Ko}}, \bibinfo {author} {\bibfnamefont
  {B.~S.}\ \bibnamefont {Kim}}, \bibinfo {author} {\bibfnamefont {B.~G.}\
  \bibnamefont {Jang}}, \bibinfo {author} {\bibfnamefont {J.~M.}\ \bibnamefont
  {Ok}}, \bibinfo {author} {\bibfnamefont {J.}~\bibnamefont {Lee}}, \bibinfo
  {author} {\bibfnamefont {Y.~J.}\ \bibnamefont {Jo}}, \bibinfo {author}
  {\bibfnamefont {W.}~\bibnamefont {Kang}}, \bibinfo {author} {\bibfnamefont
  {J.~H.}\ \bibnamefont {Shim}}, \bibinfo {author} {\bibfnamefont
  {C.}~\bibnamefont {Kim}}, \bibinfo {author} {\bibfnamefont {H.~W.}\
  \bibnamefont {Yeom}}, \bibinfo {author} {\bibfnamefont {B.}~\bibnamefont
  {Il~Min}}, \bibinfo {author} {\bibfnamefont {B.-J.}\ \bibnamefont {Yang}},\
  and\ \bibinfo {author} {\bibfnamefont {J.~S.}\ \bibnamefont {Kim}},\
  }\bibfield  {title} {\bibinfo {title} {Large anomalous {Hall} current induced
  by topological nodal lines in a ferromagnetic van der {Waals} semimetal},\
  }\href {https://doi.org/10.1038/s41563-018-0132-3} {\bibfield  {journal}
  {\bibinfo  {journal} {Nat. Mater.}\ }\textbf {\bibinfo {volume} {17}},\
  \bibinfo {pages} {794} (\bibinfo {year} {2018})}\BibitemShut {NoStop}%
\bibitem [{\citenamefont {Roy}\ \emph {et~al.}(2020)\citenamefont {Roy},
  \citenamefont {Singha}, \citenamefont {Ghosh}, \citenamefont {Pariari},\ and\
  \citenamefont {Mandal}}]{Roy2020}%
  \BibitemOpen
  \bibfield  {author} {\bibinfo {author} {\bibfnamefont {S.}~\bibnamefont
  {Roy}}, \bibinfo {author} {\bibfnamefont {R.}~\bibnamefont {Singha}},
  \bibinfo {author} {\bibfnamefont {A.}~\bibnamefont {Ghosh}}, \bibinfo
  {author} {\bibfnamefont {A.}~\bibnamefont {Pariari}},\ and\ \bibinfo {author}
  {\bibfnamefont {P.}~\bibnamefont {Mandal}},\ }\bibfield  {title} {\bibinfo
  {title} {Anomalous {Hall} effect in the half-metallic {Heusler} compound
  {Co$_2$Ti$X$ ($X$ = Si, Ge)}},\ }\href
  {https://doi.org/10.1103/PhysRevB.102.085147} {\bibfield  {journal} {\bibinfo
   {journal} {Phys. Rev. B}\ }\textbf {\bibinfo {volume} {102}},\ \bibinfo
  {pages} {085147} (\bibinfo {year} {2020})}\BibitemShut {NoStop}%
\bibitem [{\citenamefont {Gangwar}\ \emph {et~al.}(2025)\citenamefont
  {Gangwar}, \citenamefont {Tewari},\ and\ \citenamefont {Yadav}}]{Sunil2025}%
  \BibitemOpen
  \bibfield  {author} {\bibinfo {author} {\bibfnamefont {S.}~\bibnamefont
  {Gangwar}}, \bibinfo {author} {\bibfnamefont {G.~C.}\ \bibnamefont
  {Tewari}},\ and\ \bibinfo {author} {\bibfnamefont {C.~S.}\ \bibnamefont
  {Yadav}},\ }\bibfield  {title} {\bibinfo {title} {Berry curvature induced
  anomalous {Hall} and {Nernst} effects in a magnetic nodal line semimetal:
  {Mn$_3$ZnC}},\ }\href {https://doi.org/10.1103/PhysRevB.111.195106}
  {\bibfield  {journal} {\bibinfo  {journal} {Phys. Rev. B}\ }\textbf {\bibinfo
  {volume} {111}},\ \bibinfo {pages} {195106} (\bibinfo {year}
  {2025})}\BibitemShut {NoStop}%
\bibitem [{\citenamefont {Tian}\ \emph {et~al.}(2009)\citenamefont {Tian},
  \citenamefont {Ye},\ and\ \citenamefont {Jin}}]{Tian2009}%
  \BibitemOpen
  \bibfield  {author} {\bibinfo {author} {\bibfnamefont {Y.}~\bibnamefont
  {Tian}}, \bibinfo {author} {\bibfnamefont {L.}~\bibnamefont {Ye}},\ and\
  \bibinfo {author} {\bibfnamefont {X.}~\bibnamefont {Jin}},\ }\bibfield
  {title} {\bibinfo {title} {Proper scaling of the anomalous {Hall} effect},\
  }\href {https://doi.org/10.1103/PhysRevLett.103.087206} {\bibfield  {journal}
  {\bibinfo  {journal} {Phys. Rev. Lett.}\ }\textbf {\bibinfo {volume} {103}},\
  \bibinfo {pages} {087206} (\bibinfo {year} {2009})}\BibitemShut {NoStop}%
\bibitem [{\citenamefont {Shen}\ \emph {et~al.}(2020)\citenamefont {Shen},
  \citenamefont {Yao}, \citenamefont {Zeng}, \citenamefont {Sun}, \citenamefont
  {Xi}, \citenamefont {Wu}, \citenamefont {Wang}, \citenamefont {Shen},
  \citenamefont {Liu},\ and\ \citenamefont {Liu}}]{Shen2020}%
  \BibitemOpen
  \bibfield  {author} {\bibinfo {author} {\bibfnamefont {J.}~\bibnamefont
  {Shen}}, \bibinfo {author} {\bibfnamefont {Q.}~\bibnamefont {Yao}}, \bibinfo
  {author} {\bibfnamefont {Q.}~\bibnamefont {Zeng}}, \bibinfo {author}
  {\bibfnamefont {H.}~\bibnamefont {Sun}}, \bibinfo {author} {\bibfnamefont
  {X.}~\bibnamefont {Xi}}, \bibinfo {author} {\bibfnamefont {G.}~\bibnamefont
  {Wu}}, \bibinfo {author} {\bibfnamefont {W.}~\bibnamefont {Wang}}, \bibinfo
  {author} {\bibfnamefont {B.}~\bibnamefont {Shen}}, \bibinfo {author}
  {\bibfnamefont {Q.}~\bibnamefont {Liu}},\ and\ \bibinfo {author}
  {\bibfnamefont {E.}~\bibnamefont {Liu}},\ }\bibfield  {title} {\bibinfo
  {title} {Local disorder-induced elevation of intrinsic anomalous {Hall}
  conductance in an electron-doped magnetic {Weyl} semimetal},\ }\href
  {https://doi.org/10.1103/PhysRevLett.125.086602} {\bibfield  {journal}
  {\bibinfo  {journal} {Phys. Rev. Lett.}\ }\textbf {\bibinfo {volume} {125}},\
  \bibinfo {pages} {086602} (\bibinfo {year} {2020})}\BibitemShut {NoStop}%
\bibitem [{\citenamefont {Chen}\ \emph {et~al.}(2011)\citenamefont {Chen},
  \citenamefont {Shi}, \citenamefont {Xu}, \citenamefont {Zhang}, \citenamefont
  {Du},\ and\ \citenamefont {Zhou}}]{Chen2011}%
  \BibitemOpen
  \bibfield  {author} {\bibinfo {author} {\bibfnamefont {M.}~\bibnamefont
  {Chen}}, \bibinfo {author} {\bibfnamefont {Z.}~\bibnamefont {Shi}}, \bibinfo
  {author} {\bibfnamefont {W.~J.}\ \bibnamefont {Xu}}, \bibinfo {author}
  {\bibfnamefont {X.~X.}\ \bibnamefont {Zhang}}, \bibinfo {author}
  {\bibfnamefont {J.}~\bibnamefont {Du}},\ and\ \bibinfo {author}
  {\bibfnamefont {S.~M.}\ \bibnamefont {Zhou}},\ }\bibfield  {title} {\bibinfo
  {title} {Tuning anomalous {Hall} conductivity in $\textit{L}1_0$ {FePt} films
  by long range chemical ordering},\ }\href {https://doi.org/10.1063/1.3556616}
  {\bibfield  {journal} {\bibinfo  {journal} {Appl. Phys. Lett.}\ }\textbf
  {\bibinfo {volume} {98}},\ \bibinfo {pages} {082503} (\bibinfo {year}
  {2011})}\BibitemShut {NoStop}%
\bibitem [{\citenamefont {Zhu}\ \emph {et~al.}(2014)\citenamefont {Zhu},
  \citenamefont {Pan},\ and\ \citenamefont {Zhao}}]{Zhu2014}%
  \BibitemOpen
  \bibfield  {author} {\bibinfo {author} {\bibfnamefont {L.~J.}\ \bibnamefont
  {Zhu}}, \bibinfo {author} {\bibfnamefont {D.}~\bibnamefont {Pan}},\ and\
  \bibinfo {author} {\bibfnamefont {J.~H.}\ \bibnamefont {Zhao}},\ }\bibfield
  {title} {\bibinfo {title} {Anomalous {Hall} effect in epitaxial
  {$\textit{L}$1$_0$-Mn$_{1.5}$Ga} films with variable chemical ordering},\
  }\href {https://doi.org/10.1103/PhysRevB.89.220406} {\bibfield  {journal}
  {\bibinfo  {journal} {Phys. Rev. B}\ }\textbf {\bibinfo {volume} {89}},\
  \bibinfo {pages} {220406(R)} (\bibinfo {year} {2014})}\BibitemShut {NoStop}%
\bibitem [{\citenamefont {Zhu}\ \emph {et~al.}(2012)\citenamefont {Zhu},
  \citenamefont {Nie}, \citenamefont {Meng}, \citenamefont {Pan}, \citenamefont
  {Zhao},\ and\ \citenamefont {Zheng}}]{Zhu2012}%
  \BibitemOpen
  \bibfield  {author} {\bibinfo {author} {\bibfnamefont {L.}~\bibnamefont
  {Zhu}}, \bibinfo {author} {\bibfnamefont {S.}~\bibnamefont {Nie}}, \bibinfo
  {author} {\bibfnamefont {K.}~\bibnamefont {Meng}}, \bibinfo {author}
  {\bibfnamefont {D.}~\bibnamefont {Pan}}, \bibinfo {author} {\bibfnamefont
  {J.}~\bibnamefont {Zhao}},\ and\ \bibinfo {author} {\bibfnamefont
  {H.}~\bibnamefont {Zheng}},\ }\bibfield  {title} {\bibinfo {title}
  {Multifunctional \textit{{L}}1$_{\textrm{0}}$‐{Mn}$_{\textrm{1.5}}${Ga}
  films with ultrahigh coercivity, giant perpendicular magnetocrystalline
  anisotropy and large magnetic energy product},\ }\href
  {https://doi.org/10.1002/adma.201200805} {\bibfield  {journal} {\bibinfo
  {journal} {Adv. Mater.}\ }\textbf {\bibinfo {volume} {24}},\ \bibinfo {pages}
  {4547} (\bibinfo {year} {2012})}\BibitemShut {NoStop}%
\bibitem [{\citenamefont {Zhu}\ \emph {et~al.}(2013)\citenamefont {Zhu},
  \citenamefont {Pan}, \citenamefont {Nie}, \citenamefont {Lu},\ and\
  \citenamefont {Zhao}}]{Zhu2013}%
  \BibitemOpen
  \bibfield  {author} {\bibinfo {author} {\bibfnamefont {L.~J.}\ \bibnamefont
  {Zhu}}, \bibinfo {author} {\bibfnamefont {D.}~\bibnamefont {Pan}}, \bibinfo
  {author} {\bibfnamefont {S.~H.}\ \bibnamefont {Nie}}, \bibinfo {author}
  {\bibfnamefont {J.}~\bibnamefont {Lu}},\ and\ \bibinfo {author}
  {\bibfnamefont {J.~H.}\ \bibnamefont {Zhao}},\ }\bibfield  {title} {\bibinfo
  {title} {Tailoring magnetism of multifunctional {Mn}$_{x}${Ga} films with
  giant perpendicular anisotropy},\ }\href {https://doi.org/10.1063/1.4799344}
  {\bibfield  {journal} {\bibinfo  {journal} {Appl. Phys. Lett.}\ }\textbf
  {\bibinfo {volume} {102}},\ \bibinfo {pages} {132403} (\bibinfo {year}
  {2013})}\BibitemShut {NoStop}%
\bibitem [{\citenamefont {Vilanova~Vidal}\ \emph {et~al.}(2011)\citenamefont
  {Vilanova~Vidal}, \citenamefont {Schneider},\ and\ \citenamefont
  {Jakob}}]{Vidal2011}%
  \BibitemOpen
  \bibfield  {author} {\bibinfo {author} {\bibfnamefont {E.}~\bibnamefont
  {Vilanova~Vidal}}, \bibinfo {author} {\bibfnamefont {H.}~\bibnamefont
  {Schneider}},\ and\ \bibinfo {author} {\bibfnamefont {G.}~\bibnamefont
  {Jakob}},\ }\bibfield  {title} {\bibinfo {title} {Influence of disorder on
  anomalous {Hall} effect for {Heusler} compounds},\ }\href
  {https://doi.org/10.1103/PhysRevB.83.174410} {\bibfield  {journal} {\bibinfo
  {journal} {Phys. Rev. B}\ }\textbf {\bibinfo {volume} {83}},\ \bibinfo
  {pages} {174410} (\bibinfo {year} {2011})}\BibitemShut {NoStop}%
\bibitem [{\citenamefont {He}\ \emph {et~al.}(2012)\citenamefont {He},
  \citenamefont {Ma}, \citenamefont {Shi}, \citenamefont {Guo}, \citenamefont
  {Zheng}, \citenamefont {Xin},\ and\ \citenamefont {Zhou}}]{He2012}%
  \BibitemOpen
  \bibfield  {author} {\bibinfo {author} {\bibfnamefont {P.}~\bibnamefont
  {He}}, \bibinfo {author} {\bibfnamefont {L.}~\bibnamefont {Ma}}, \bibinfo
  {author} {\bibfnamefont {Z.}~\bibnamefont {Shi}}, \bibinfo {author}
  {\bibfnamefont {G.~Y.}\ \bibnamefont {Guo}}, \bibinfo {author} {\bibfnamefont
  {J.~G.}\ \bibnamefont {Zheng}}, \bibinfo {author} {\bibfnamefont
  {Y.}~\bibnamefont {Xin}},\ and\ \bibinfo {author} {\bibfnamefont {S.~M.}\
  \bibnamefont {Zhou}},\ }\bibfield  {title} {\bibinfo {title} {Chemical
  composition tuning of the anomalous {Hall} effect in isoelectronic
  $\textit{L}1_0$ {FePd$_{1-x}$Pt$_x$} alloy films},\ }\href
  {https://doi.org/10.1103/PhysRevLett.109.066402} {\bibfield  {journal}
  {\bibinfo  {journal} {Phys. Rev. Lett.}\ }\textbf {\bibinfo {volume} {109}},\
  \bibinfo {pages} {066402} (\bibinfo {year} {2012})}\BibitemShut {NoStop}%
\end{thebibliography}%
%\end{footnotesize}

\end{document}